\documentclass[12pt]{article}

\usepackage[utf8]{inputenc}
\DeclareUnicodeCharacter{0306}{\u}

\usepackage{graphicx}
\usepackage{xcolor}
\usepackage[normalem]{ulem}
\usepackage{amsmath}
\usepackage{float}
\usepackage{tabularx,booktabs}
\usepackage{bm}
\usepackage{comment}
\usepackage{soul}
\usepackage{lmodern}
\usepackage{setspace}
\usepackage{vicent}
\usepackage[OT1]{fontenc}
\usepackage{mathtools}
\usepackage{multicol}
\usepackage{tcolorbox}
\usepackage{nameref}
\usepackage{textcomp}
\usepackage{url}
\usepackage[breaklinks]{hyperref}
\usepackage{bm}

\newsavebox{\boxedmaterialbox}

\newcommand{\cblu}{\color{blue}}
\newcommand{\cred}{\color{red}}

\newcommand{\pid}{$p_{ID}$}
\newcommand{\pinc}{$p_{Inc}$}
\newcommand{\pex}{$p_{Exc}$}
\newcommand{\qid}{$q_{ID}$}
\newcommand{\qinc}{$q_{Inc}$}
\newcommand{\qex}{$q_{Exc}$}

\newcommand{\myfrac}[2]{\displaystyle{\frac{#1}{#2}}}

\usepackage{color}
\newcount\Comments  % 0 suppresses notes to selves in text
\definecolor{darkgreen}{rgb}{0,0.5,0}
\definecolor{darkred}{rgb}{0.7,0,0}
\definecolor{teal}{rgb}{0.1,0.6,0.7}
\definecolor{blue}{rgb}{0.0,0.1,0.9}
\definecolor{orange}{rgb}{1.,0.7,0.0}
\definecolor{lightblue}{rgb}{0.70, 0.80, 0.89}
\definecolor{mediumblue}{rgb}{0.60, 0.70, 1}
\definecolor{pink}{rgb}{0.3,0,0}
\definecolor{yellow}{rgb}{0,0.5,0.5}

\newcommand{\kibitz}[2]{\ifnum\Comments=1{{\textcolor{#1}{\textsf{\footnotesize [#2]}}}}\fi}

\newcommand{\hari}[1]{\kibitz{darkred}{Hari: #1}}

\usepackage{url}
\usepackage{dirtytalk}

\usepackage[style=apa, citestyle=authoryear, backend=biber,natbib, uniquename=false]{biblatex}
\newcolumntype{Y}{>{\centering\arraybackslash}X}

\makeatletter
\NewDocumentCommand{\sotwo}{O{blue}O{black}+m}
    {%
        \begingroup
        \color{#1}%
        \setul{-.5ex}{.4pt}%
        \def\SOUL@uleverysyllable{%
            \rlap{%
                \color{#2}\the\SOUL@syllable
                \SOUL@setkern\SOUL@charkern}%
            \SOUL@ulunderline{%
                \phantom{\the\SOUL@syllable}}%
        }%
        \ul{#3}%
        \endgroup
    }
\makeatother

\begin{document}

%\centerline{\Large  \bf The Role of Bayesian Reasoning} 
%\centerline{\Large \bf Bayesian Reasoning and the Importance of Validation Data}

\centerline{\Large \bf The Influence of Validation Data on Logical and Scientific } 
\vskip 0.1in
\centerline{\Large \bf Interpretations of Forensic Expert Opinions }

\vskip 0.2in
\centerline{Steven Lund (steven.lund@nist.gov) and Hari Iyer (hari@nist.gov)}
\centerline{Statistical Engineering Division, ITL/NIST, Gaithersburg, MD 20899}
\centerline{Contact author email: steven.lund@nist.gov}
\doublespacing
\normalsize
\newpage
\begin{abstract}

Forensic experts use specialized training and knowledge to enable other members of the judicial system to make better informed and more just decisions. Factfinders, in particular, are tasked with judging how much weight to give to experts' reports and opinions.  Many references describe assessing evidential weight from the perspective of a forensic expert. Some recognize that stakeholders are each responsible for evaluating their own weight of evidence.  Morris (1971, 1974, 1977) provided a general framework for recipients to update their own uncertainties after learning an expert's opinion.  Although this framework is normative under Bayesian axioms and several forensic scholars advocate the use of Bayesian reasoning, few resources describe its application in forensic science.  This paper addresses this gap by examining how recipients can combine principles of science and Bayesian reasoning to evaluate their own likelihood ratios for expert opinions. This exercise helps clarify how an expert's role depends on whether one envisions recipients to be logical and scientific or deferential.  Illustrative examples with an expert's opinion expressed as a categorical conclusion, likelihood ratio, or range of likelihood ratios, or with likelihood ratios from multiple experts, each reveal the importance and influence of validation data for logical recipients' interpretations.

\end{abstract}

\newpage

\section{Introduction}
\label{sec:intro}

Much has been written about the role of Bayesian reasoning in assessing weight of forensic evidence, most commonly considering how experts could, or should, summarize their findings using likelihood ratios ($LR$s) (e.g., see \cite{lindley1977problem,  aitken2004statistics,   
aitken2008fundamentals, aitken2010fundamentals, robertson2016interpreting,fienberg1983presentation}).  In many judicial systems around the world, however, it is the non-experts, such as jurors, magistrates, or judges, who are tasked with assessing how much weight to give to the information or evidence presented by a forensic expert.   While forensic experts interpret physical or digital evidence, other stakeholders are often tasked with interpreting the resulting expert opinions.  

The problem of logically interpreting someone else's opinion has been considered by many Bayesian experts (e.g., see \cite{morris1977combining,  french1980updating,genest1985modeling,ouchi2004literature,Aspinall2013QuantifyingSU}). In \citet{de2017theory}, this challenge is described as ``subjective squared:  our subjective judgment regarding the subjective judgment of others."    

To understand and demonstrate how well their methods perform in analyzing and interpreting evidence, forensic experts often participate in validation testing, where the methods and procedures used in casework are applied to ground-truth known examples.  Performance data from validation tests can help stakeholders interpret expert opinions. In some of the papers most directly relevant to forensic practice, Peter Morris laid out a comprehensive and logical framework to interpret expert opinions, including when performance data is available (\cite{morris1971bayesian}; \cite{morris1974decision}; \cite{morris1977combining}).  Morris' framework is normative if one accepts the axioms underlying Bayesian reasoning.

In particular, \citet{morris1977combining} provided a detailed description of the process for a decision maker to update their uncertainty after learning expert opinions along with corresponding performance data.  The framework allows for multiple expert opinions, even conflicting ones. Morris illustrated the process in scenarios where experts provide their opinions in the form of their probability distributions describing their state of uncertainty regarding an unknown  quantity of interest. However, the framework can be applied regardless of the form that expert opinion takes, whether it is a categorical conclusion, a posterior probability, an $LR$, or a range of $LR$ values.  Even if the expert simply reports the output of a computer algorithm - it is all just `new information' to the decision maker.  In any case, each decision maker proceeds by assessing their likelihood ratio for this new information. This is done by assigning the probability of the new information under each  proposition (e.g., $H_1$ and $H_2$) of interest to the decision maker.   Morris further described the role of performance or validation data in shaping these probabilities.  

The process of updating uncertainties, once an initial state of uncertainty has been fully specified in terms of probability distributions, is simply a computational exercise.  However, selecting a specific distribution to represent one's initial state of uncertainty is a challenging task and any given choice may seem somewhat arbitrary, even in simple situations. 
%Though conceptually straightforward, Bayesian reasoning is rarely straightforward to implement in practice because in most real world applications one has to express one's initial uncertainties in terms of joint distributions of a number of unknown quantities. There often is no systematic approach to accomplish this task and many of the choices seem quite arbitrary. 
This may explain why few, if any, writings by forensic scholars reflect or promote the normative process for decision makers interpreting expert opinions, despite the fact that many forensic scholars endorse the Bayesian paradigm and recognize it as normative for the decision maker.

Consider two basic takeaways from the  framework described in \citet{morris1977combining}.

\noindent{\bf Takeaway 1: }
Recipients must interpret expert opinions for themselves. 

Discussions of logical or Bayesian reasoning in forensics appear divided on this point. While some scholars acknowledge that logic requires stakeholders to assign their own weight to forensic evidence or an expert's opinion (e.g., see \cite{lindley2013understanding, gittelson2018response,fienberg1996bayesian, evett2015logical,fienberg1983presentation}), others envision a system where stakeholders defer to an expert's $LR$ assessment by accepting it as their own (e.g., see \cite{ aitken2008fundamentals,  BIEDERMANN2017, biedermann2018analysing,  robertson2016interpreting}).

\noindent{\bf Takeaway 2:}
Validation data helps shape decision maker's interpretations of expert opinions. 

Few, if any, writings by forensic scholars reflect the normative role of validation data in a decision maker's uncertainty.  The legal and forensic community often consider validation in a binary manner.  (This often prompts questions like "has this method been validated?" or "how many tests are needed to validate this method?")    Saying an expert used a validated method to arrive at their opinion seems like an assurance that the risks of being substantially misled by the expert's opinion are low enough to justify decision makers accepting the expert's opinion as their own. Under the normative approach of Bayesian reasoning, summarizing validation data in this manner loses substantial information, such as how many tests were conducted, under what conditions, and what results were obtained.  %{\cred From Will: Does treating validation data in this binary manner really risk just a loss of information that presumably cuts both ways, or could it be said to lead to systematic mischaracterization of a decision maker's interpretation of the information?    More generally, is it possible to say something about the cost of the information loss caused by treating validation in a binary manner?}
%In particular, validation data is valuable and influential before for  can be as important to the decision maker as the expert's opinion. 

%Stakeholders deferring to expert opinions simplifies life by skirting the complexities of de Finetti's subjectivity squared challenge, but the trade-off is potentially weakening safeguards against unjust decisions. M. Chris Fabricant’s recent book “Junk Science and the American Criminal Justice System” \parencite{fabricant2022junk} details several instances of wrongful convictions.  A common thread is that the expert's confidence was not checked against empirical tests of the expert's performance.

Instead of emphasizing either of the above takeaways, forensic scholars discussing logical or Bayesian inference often argue that experts should provide their opinions as likelihood ratios.  This restriction does not follow from the normative approach discussed in \citet{morris1977combining} for the decision maker, since that approach is agnostic to the form of the opinion provided by the expert. That is, there is nothing normative or logical about requiring an expert to provide their opinion in the form of a $LR$. Of course, the expert is free to choose what they think is the best way to communicate the value of evidence to the decision maker.

In this paper, we aim to increase awareness and understanding among the forensic and legal communities of the logical approach for stakeholders assigning weight to forensic expert opinions, including the logical role of validation data.  %we consider the application of Bayesian reasoning \parencite{morris1977combining} by the decision maker for interpreting forensic expert opinions, especially when accompanied by their performance data.  %We note that there is no need to restrict expert opinions to be their personal probabilities. 
We illustrate the process using examples readily recognizable and relevant to the forensic science community. In particular, we  provide illustrations of recipients interpreting expert opinions when expressed as a categorical conclusion, a likelihood ratio, or a range of likelihood ratios.\footnote{Though not directly illustrated, the processes described in this paper also apply to more ``objective'' results reported by an expert, such as a similarity score output by a computer algorithm.} %We hope to contribute to conversations regarding how validation data is gathered and, ultimately, presented alongside expert opinions.

We believe this topic is relevant for a wide range of stakeholders, including experts and lawyers, who regularly interact with testimony and reports from forensic experts.  An expert's ultimate act in a particular case is to communicate their findings and opinions to others. Envisioning the subsequent step in which recipients assess what weight to give an expert's testimony or report leads to important questions regarding what and how an expert should communicate.  Is the information presented accurate, and would recipients find it both understandable and helpful?  After hearing from the expert, do recipients understand it is their responsibility, and theirs alone, to assess  the weight of the expert's opinion or conclusion, or do they feel the expert's perspective is intended to be communal (i.e., the recipients should adopt the expert’s opinion)?  What information can recipients use to assess what weight  %(or reliability) of the 
to give to an expert's opinion or conclusion in the case at hand?

The remainder of the paper is organized as follows.  Section~2 is devoted to a brief discussion of logical and scientific reasoning and the ‘personal’ nature of probabilities.  Section~3 contains several examples of a hypothetical fact finder using logical and scientific reasoning to assign a weight to an expert's opinion. Examples include an expert providing a categorical conclusion, a  $LR$, and a $LR$ range, or even a situation where two experts each provide their own $LR$. Section~4 is devoted to a summary discussion and concluding remarks.

\section{Logical and Scientific  Reasoning}
\subsection{Logical Reasoning}

As with many previous writings (e.g., \cite{de2017theory,lindley2013understanding,kadane2020principles,evett2015logical,buckleton2018forensic}), we consider logical reasoning to be synonymous with abiding by the laws of probability. In particular, we require conformance with Bayes rule when handling uncertainty in the context of interpreting expert opinions as in \citet{morris1977combining}.  That is, any reasoning found to violate Bayes rule is considered illogical.  We now provide a brief overview of how Bayes rule provides a logical approach to update one's uncertainty in response to new information.

%In the simplest application, where someone is trying to decide which of two mutually exclusive and exhaustive propositions is true, assessing whether or not an individual's reasoning violated Bayes rule requires three quantities:  their prior odds, their $LR$, and  their posterior odds. These are discussed in greater detail in Sections~\ref{sec:ScienceAndLogic} and \ref{sec:br}. 

%\subsubsection{Bayesian Reasoning}
%\label{sec:br}
Bayesian reasoning refers to the practice of evaluating and updating uncertainties in a manner that conforms with Bayes' rule, which describes the constraints the laws of probability place on how an individual's beliefs should be affected by new information. More specifically, it shows how three probabilistic quantities, namely prior odds, likelihood ratios (or Bayes' factors), and posterior odds, must be related to one another in order to comply with the laws of probability. In its simplest application, Bayes' rule would apply to a person updating their belief regarding which of two propositions, say $H_1$ and $H_2$, is true, in light of some newly encountered information.  For instance, $H_1$ may reflect a perspective corresponding to the prosecution, and $H_2$ may reflect a perspective corresponding to the defense.\footnote{In order for an individual using Bayes’ rule to arrive at valid updated or “posterior” probabilities of the propositions given the presented evidence, the individual must consider an exhaustive set of propositions. That is, any proposition for which a decision maker has a non-zero prior probability must be included in the set of propositions considered by that decision maker.
When there are more than two mutually exclusive propositions, Bayes' rule can still be used to obtain posterior probabilities as a function of prior probabilities and likelihoods associated with each proposition. For ease of discussion, and without loss of generality, we restrict our presentations to the simplest scenario of two mutually exclusive and exhaustive propositions.}

Suppose an individual characterizes how sure they feel about the truth of $H_1$ using probability $p_1$ and about the truth of $H_2$ using probability $p_2$.  The ratio of these two probabilities, $O_{12}=\myfrac{p_1}{p_2}$, is often called the odds  of $H_1$ versus $H_2$.  Upon encountering new information, say $E$, the individual's uncertainties regarding the truth of $H_1$ and $H_2$ may change. Thus, there is a distinction between the individual's probabilities as assessed before, or prior to, learning $E$ and the individual's probabilities as assessed after, or posterior to, learning $E$. The individual's original probabilities, $p_1$ and $p_2$, evaluated before encountering the new information, are referred to as prior probabilities (relative to $E$) and their ratio as prior odds of $H_1$ versus $H_2$.  Probabilities reflecting the individual's uncertainty regarding the truth of $H_1$ and $H_2$ after learning the new information, say $p^*_1$ and $p^*_2$, respectively, are referred to as posterior probabilities (relative to $E$), and the ratio $O^*_{12}=\myfrac{p^*_1}{p^*_2}$ is referred to as posterior odds of $H_1$ versus $H_2$.  The impact of the new information $E$ on the individual's uncertainty is reflected by the ratio $\myfrac{O^*_{12}}{O_{12}}$.  For instance, if the prior and posterior odds of $H_1$ versus $H_2$ are very similar, their ratio will be close to 1, indicating that the new information has had very little impact on the individual's uncertainties.   

Bayes' rule relates the ratio of posterior and prior odds of $H_1$ versus $H_2$ to what is known as a likelihood ratio.  A $LR$ for $E$ is computed as the ratio between how likely one feels $E$ would be to occur if $H_1$ were true, say $l_1$, and how likely one feels $E$ would be to occur if $H_2$ were true, say $l_2$.  Bayes' rule requires  the ratio of an individual's posterior odds to their prior odds, $\myfrac{O^*_{12}}{O_{12}}$, to be equal to the individual's likelihood ratio, $LR_{12}=\myfrac{l_1}{l_2}$.  This can be equivalently restated as a requirement that an individual's posterior odds  are equal to the product of the individual's corresponding prior odds and the individual's likelihood ratio. Any triplet of prior odds, likelihood ratio, and posterior odds that fail to conform to Bayes rule violates the basic laws of probability theory and may be labeled as illogical, irrational, or incoherent.

Thus, at least in theory, Bayes' rule provides a pathway for how a logical person could update their uncertainty, after encountering new information in three steps: (1) Assess how sure you feel about the truth of each considered proposition;  (2) Assess how likely the newly encountered information would be to have occurred, assuming the truth of each considered proposition in turn;  (3) Compute posterior probabilities in accordance with Bayes' rule. This general process is the same, regardless of what form the new information takes.  We say ``at least in theory'' because assessing probabilities is rarely as straightforward as it might seem. 

The viewpoints expressed in this paper are natural consequences of accepting that uncertainty is personal. Although rules of probability may seem rigorous and exact, they are only ``if-then'' statements as in, ``if your prior odds are 0.1 and your likelihood ratio is 100, then your posterior odds must be 10."   The laws of probability do not tell you what priors to start with or what likelihoods to use, so they do not dictate the appropriate probabilistic interpretation of any given situation.  Consequently, probabilistic assessments can be expected to vary from one logical person to another. For the convenience of readers who may feel uncomfortable or unfamiliar with the perspective that probabilities are personal, and consequently judgments of strengths of evidence are personal, we provide additional discussion and an illustrative example in Appendix-A.

 The personal nature of probability has been explicitly emphasized by scholars in probability and statistics (e.g., \cite{lindley2013understanding, kadane2020principles, de2017theory}) and in forensics (e.g.,  \cite{champod2009evidence,biedermann2013your,taroni2016dismissal, berger2016lr, gittelson2018response}). This makes it important to specify whose probabilities are being discussed in applications involving multiple people processing information and assigning probabilities. Lindley's use of the word ``your'' when addressing a hypothetical juror in the following excerpt emphasizes this point. ``We saw in §6.6 how evidence $E$ before the court would change your probability to $p(G|E K)$  using Bayes’ rule. The calculation required by the rule needs your likelihood ratio  $p(E|G K)/p(E|G^c K)$, involving your probabilities of the evidence, both under the supposition of guilt and of innocence...''\footnote{ Lindley uses $G$ and $G^c$ to stand for guilty and not-guilty, respectively.  These two mutually exclusive and exhaustive propositions are represented as $H_1$ and $H_2$ in our notation.  Lindley uses $K$ to represent the juror's background information.}(page 260, Section 10.14 titled ``Legal Applications", \textcite{lindley2013understanding})   %\st{Importantly, Lindley did not suggest the juror's $LR$ would match the expert's or that the juror should use the expert's $LR$ when applying Bayes' formula.}

Throughout this paper, we will specify who has assigned the value for each probability.  While somewhat tedious, we do this to consistently acknowledge that there are multiple individuals with a responsibility to consider forensic evidence and that probabilities are personal.  
%\subsection{Applying Bayes' Rule to Expert Opinions}
%\section{Applying Bayes' Rule to Expert Opinions}
In particular, this leads to separate instances of Bayes rule for the recipient and the expert.
 
Consider Bayes rule for the recipient, which is given by
\begin{equation}
\mbox{Posterior Odds}_{Rational Recipient} = \mbox{Prior Odds}_{Rational Recipient} \times \mbox{LR}_{Rational Recipient}.
\label{eq:BayesRational}
\end{equation}
All components in this expression belong to the recipient.  An expert's opinion (and other information they may provide) serves as the input to the recipient's $LR$.  Unless a recipient is deferential, the probabilities in Bayes rule for the recipient are distinct from those in Bayes rule for the expert, which would be given by
\begin{equation}
\mbox{Posterior Odds}_{Expert} = \mbox{Prior Odds}_{Expert} \times \mbox{LR}_{Expert}.
\label{eq:BayesExpert}
\end{equation}

Notwithstanding the recognition by the community that probabilities are personal, when Bayes' rule is brought up in forensic contexts, it is often presented in general terms that do not specify to whom the probabilities belong.   %Some forensic scholars, (for instance, \cite{aitken2018roles}), say it is ``perfectly appropriate'' for recipients of an expert's $LR$ to combine their own prior odds with the expert's likelihood ratio to obtain their posterior odds in favor of $H_1$ compared to $H_2$. 
Although we are not aware of anyone explicitly arguing that decision makers should not use their own personal $LR$s to arrive at their posterior odds,  weight of forensic evidence discussions often fail to convey clearly that each party has their own Bayes' equation and emphasize that an expert's role is provide a $LR$. 
We believe such discussions indirectly encourage decision makers to use a $LR$ provided by an expert in place of their own.  This corresponds to what we refer to as the  ``Deferential Bayes" equation, \begin{equation}
\mbox{Posterior Odds}_{Deferential Recipient} = \mbox{Prior Odds}_{Deferential Recipient} \times \mbox{LR}_{Expert}.
\label{eq:BayesHybrid}
\end{equation}
Although it attempts to leave assessments of prior odds (and costs of errant decisions) to the recipient, this scenario represents a departure from Bayesian reasoning \parencite{lund2017likelihood}.   %As we mentioned earlier, some experts explicitly endorse this equation \parencite{robertson1992unhelpful,biedermann2014liberties, biedermann2018analysing, taroni2016dismissal}. Other LR proponents, however, acknowledge that the recipient is responsible for assessing their own LR \parencite{lindley2013understanding, gittelson2018response}. 
Appendix-B provides excerpts from the forensic science literature that either explicitly endorse or appear to indirectly support the use of the deferential-Bayes equation.

It is worth mentioning that logic does not prevent a recipient from waiting to form any probabilistic assessments until after encountering all relevant information  (See \citet{lindley2013understanding,good1991weight}).  That is, going straight to the posterior without employing Bayes Rule is not illogical. In the illustrative examples in Section~\ref{sec:bayesianrecipients}, we choose to specify prior distributions and likelihoods so that the influence of the expert's opinion on the recipient's uncertainty is explicitly articulated. This also helps clarify the effect validation data has on the weight our example recipient gives to the expert's opinion.

It is also worth mentioning that Bayesian reasoning does not restrict what type of information could influence the weight a recipient gives to an expert's opinion.  We hope readers agree it would be undesirable for things like an expert's accent, clothing, or body language to affect the outcome of a case.  For this reason, it is also important to appeal to principles of scientific reasoning rather than  Bayesian reasoning alone.

\subsection{Scientific Reasoning}
Scientific reasoning can be difficult to define, so we use the following quotes as motivation.

“It doesn’t matter how beautiful your theory is, it doesn’t matter how smart you are. If it disagrees with experiment, it’s wrong. In that simple statement is the key to science.” – Richard Feynman, Nobel Laureate, lecture on the scientific method. (See \citet{Feynman1964}.)

“In God we trust, all others bring data.” – Edward Deming  (See \citet{WEdwardsDeming}.)

The quotes above emphasize that scientific reasoning requires using data and experimentation to shape perspectives and evaluate theories. This is not limited to experts analyzing evidence in a lab. Across the judicial system, a critical recurring question is how much weight to give an expert's opinion.

The following quotes remind one that the confidence with which an expert communicates an opinion, their general level of scientific prestige, or their years of study in a particular field are not indicators of how much weight one should give to those opinions. 

“The confidence people have in their belief is not a measure of the quality of evidence but of the coherence of the story the mind has managed to construct.” – Daniel Kahneman, Nobel Laureate. (See \citet{kahneman2011thinking}.)

``Science is the organized skepticism in the reliability of expert opinion'' - Richard Feynman, Nobel Laureate. (See \citet{feynman2015quotable}.)

This is not to dismiss the value of expertise, but rather to focus where the value of expertise lies.  In particular, stakeholders rely on forensic experts to be aware of and able to apply the most effective analytic methods and to be familiar with available data that can demonstrate how effectively their chosen methods work in various situations.  While Bayesian reasoning does not preclude an expert's confidence from influencing a recipient, scientific reasoning compels a recipient to always begin with a level of skepticism and to consider the expert's  performance data from similar scenarios when assessing how much weight to give an expert's opinion. In essence, scientific reasoning says not to put much faith in experts without data.  We illustrate the influence of validation data of strength of evidence assessments in the following two examples.

\noindent {\bf Example 1:} Consider two experts giving the same statement but with substantially differing performances in previous validation tests. %{\cred Some concerns with the next sentence -- I now forgot what!!} 
Suppose one expert has conducted many ground truth known case-like evaluations with no errors, and the other expert has relatively few ground truth known case-like evaluations with several errors.   We hope advocates of science would agree that the first expert has demonstrated greater reliability, and, therefore, opinions from the first expert deserve  greater weight than opinions from the second.  Thus, information regarding the demonstrated performance can be of great value to the recipients in making judgments regarding  reliability of opinions of different experts and how much weight to give  an expert's opinion in a given case.

\noindent{\bf Example 2:} Suppose a cartridge case was recovered at a crime scene. Subsequently, a gun was recovered from a person of interest. Upon comparing cartridge cases from test firings from this firearm with the case from the crime scene, the forensic expert reaches an opinion of ``inconclusive''.
%\parencite{baldwin2014study}

On its own, an ``inconclusive'' opinion may seem like the expert has suggested that the evidence does not have relevant information that would sway someone's uncertainty regarding the source of the recovered cartridge case.  However, the black box study \cite{baldwin2014study} examined the rate at which examiners reached various conclusions when comparing cartridge cases fired by the same firearm and when comparing cartridge cases fired by different firearms of the same model.  Their results showed 11 ``inconclusive'' conclusions out of 1090 mated comparison conclusions (~1\%) and 735 ``inconclusive'' conclusions out of 2180 nonmated comparison conclusions (~33.7\%).\footnote{For simplicity, this analysis considers the data as summarized on pages 15 and 16 in \cite{baldwin2014study}.  Evaluating data at this level does not consider the three subcategories of ``inconclusive'' conclusions available in the AFTE conclusion scale.  It also ignores the effect of laboratory policies that dictate when an examiner is allowed to reach an elimination conclusion and whether an examiner is allowed to use the inconclusive subcategories. These considerations are mentioned on pages 6 and 7 of \cite{baldwin2014study}.} Comparing these rates of ``inconclusive'' conclusions leads to  a conclusion rate ratio  of (11/1090)/(735/2180) $\approx \displaystyle{\frac{1}{33}}$. 
 (\citet{guyll2023validity} describe such ratios of relative frequencies as an estimated probative value of the opinion.  One could regard this as an empirical analog of a $LR$.)   This observation may lead the recipient to disregard an expert's assessment of the evidence as ``inconclusive'' and view it as supporting an exculpatory conclusion.
Thus, focusing on performance data can help a recipient to look past the words an expert has used, when necessary, and instead more fully understand the true  meaning of the information provided by the expert from the performance data itself.

In Section~\ref{sec:bayesianrecipients}, our hypothetical recipient illustrates scientific reasoning by acknowledging a substantial, initial uncertainty regarding what opinions or interpretations a forensic expert would tend to produce under different scenarios of interest.  This uncertainty limits the weight a recipient assigns to an expert's opinion, regardless of how strong of an opinion the expert expresses.  The recipient's uncertainty reduces in response to learning about the results of validation testing, which in turn leads the recipient to give some expert opinions stronger weight.

%Given the difficulty of making high-stake decisions in the presence of uncertainty, we examine how  recipients can use science and logic to guide their assessments of expert opinions.  %In particular, we note that a deferential approach of accepting expert opinions at face value represents neither science nor a logical processing of  uncertainties.

%When facing complex, scientific evidence that requires substantial specialized training, stakeholders often take a deferential approach and accept expert opinions at face value. This deferential approach, we argue, is neither a result of logical processing of  uncertainties, nor is it science based. 

\subsection{Science and Logic in Evidence Interpretation}
\label{sec:ScienceAndLogic}

Science and logic both play a critical role in ensuring forensic evidence leads to just outcomes. The book “Junk Science and the American Criminal Justice System” \parencite{fabricant2022junk}
%by M. Chris Fabricant 
describes several cases in which undue weight was given to faulty interpretations of forensic evidence, resulting in miscarriages of justice. “Junk science” occurs when an expert’s expressed strength of evidence differs substantially from what can be empirically supported and yet recipients accept it at face value.  Ideally, recipients would apply logical and scientific reasoning, adjusting the weight they give an expert's opinion based on how well the expert's method has been demonstrated to perform and how thoroughly it has been tested.  Ensuring recipients understand both the need and the general methods to apply such reasoning would provide an additional layer of protection from junk science infiltrating the judicial system. As noted by the  \citet{law2011expert}, however, there is a risk that recipients simply defer to an expert's opinion, treating it not as one person's opinion but as unquestioned fact.

There are two main approaches currently being used by forensic experts to convey their opinions. In most pattern disciplines (e.g., friction ridge, firearms and toolmarks, footwear and tiretread), experts summarize their findings and communicate their opinions according to a categorical conclusion scale (e.g., Identification, Inconclusive, or Exclusion). In other fields, particularly DNA, it is more common for experts to report their opinion using a continuous scale, in the form of a $LR$ value, which can range between 0 and infinity.

Many forensic scholars argue that experts should use $LR$s to convey their opinions \parencite{aitken1991use,providers2009standards,robertson2016interpreting,aitken2018roles,champod2009evidence,buckleton2018forensic}. $LR$ proponents correctly note that a forensic expert would not be aware of other evidence to which the recipient may have been exposed, so experts should not assess probabilities other than likelihoods for the evidence they evaluate \parencite{aitken1991use,providers2009standards,robertson2016interpreting,aitken2018roles,champod2009evidence,buckleton2018forensic}. Some also note that evaluating societal costs of errant decisions is not part of a forensic scientist's specialized expertise.  Thus, assessing prior odds and costs of erroneous decisions should be left to decision makers.  Since costs and prior probabilities are essential components of arriving at a decision under Bayesian reasoning,  they conclude experts should not provide categorical conclusions.  This argument is more compelling when one envisions a deferential recipient using Equation~\ref{eq:BayesHybrid} than when considering rational recipients using Equation~\ref{eq:BayesRational}. To clarify, deferential recipients believe whatever the expert says.  Rational recipients, on the other hand, only believe that the expert said it (Equation~\ref{eq:BayesExpert}) and assess for themselves what weight to give the expert's statement (Equation~\ref{eq:BayesRational}).  %The choice of an expert's opinion scale affects deferential recipients more than it affects rational recipients.

This point raises the question of what criteria could help decide what opinion scale to use.  Perhaps the most natural criterion is that the expert should use whatever method  has been shown to work best for cases like the one at hand.  In theoretical settings where the probabilities are given, $LR$s are known to lead to optimal strategies  for distinguishing between two propositions.  (In statistics, this is known as the Neyman-Pearson Lemma \parencite{neyman1933ix}.) This may inspire experts to use $LR$s to develop strongly performing methods.
In real-world applications, however, it is important to remember that $LR$s are subjective and depend on modeling choices, including the prior probabilities assigned to different propositions when more than two proposition are possible..  (These points are discussed in Appendix A, and also in \cite{lund2017likelihood}, and  \cite{berger2016lr}.)   This means that, while experts can provide $LR$s that discriminate well, they can never provide a recipient's $LR$.  

The distinction between an expert's $LR$ and a recipient's $LR$ is not the only potential point of confusion.  Many laypersons tend to confuse $LR$s as posterior odds \parencite{thompson2013jurors,thompson2015lay}. This observation reminds us of the need to consider the recipient's understanding when evaluating potential approaches to expert communication.  To that effect,  
%Some methods of communication might enable recipients to better discriminate between H1 and H2 than other methods of communication. As mentioned earlier. the obvious way  to truly judge whether one method of evidence communication is more desirable than another would be to consider which of the two approaches leads to less wrongful decisions overall. This is not directly knowable in practice. So we look for characteristics of evidence communication methods that {\em we believe} will facilitate better judicial decisions. In particular, 
we propose that evidence communication approaches should strive to be accurate, while also being understandable and useful to the recipients. By accurate, we mean that the message as articulated by the expert can be verified by authoritative sources, where experts are authorities regarding their own opinions and empirical testing results are the authority regarding performance of a given method. By understandable, we mean that the message others receive is consistent with the expert’s intended meaning and does not lead to misinterpretations that could impact the outcome of a trial. By useful, we mean that the expert anticipates a meaningful shift in the recipients’ uncertainties as a result of the information provided (i.e., the recipient's $LR$ will not be close to 1). 

The following section contains examples of a hypothetical recipient using logical and scientific reasoning to interpret expert opinions.

\section{Recipients Applying Bayesian Reasoning - General Process and Examples}
\label{sec:bayesianrecipients}

In this section, we illustrate how Alice, {\large \textbf {a l}}og{\large\textbf{i}}cal and s{\large\textbf{c}}i{\large\textbf{e}}ntific recipient, assigns weight to various forms of opinions expressed by Bert the expert. Though gifted in Bayesian reasoning, Alice is otherwise intended to represent a typical layperson without specialized knowledge regarding the methods used by forensic experts. She is scientific in the sense that she looks for relevant data when assessing uncertainties.  As a logical  person, Alice does not defer to expert opinions or interpret Bert's opinion itself as fact.  Rather, for Alice, the fact is simply that Bert stated this opinion.  %We assume Alice has correctly understood Bert's opinion.% (i.e., in the  terminology from Section X, $E_A = \phi_B(E_B,I_B)$ )
 
The examples in this section are intended as illustrations of how one could apply Morris' general approach \citep{morris1971bayesian, morris1974decision, morris1977combining} to interpret forensic expert opinions in a logical and scientific manner.  The specific choices for prior distributions and likelihoods used in these examples are placeholders rather than recommendations.  Choices of priors and likelihoods are the responsibility of each individual stakeholder. Examples include Alice interpreting Bert's opinion expressed as a categorical conclusion, a $LR$, and a $LR$ range, respectively. Alice also considers differing $LR$s for the same evidence from Bert and another expert.  Throughout the examples, we assume Alice understands the expert opinions as the experts intend. %(i.e., in the  terminology from Section X, $E_A = \phi_B(E_B,I_B)$ ) 

Each example is based on the following general setup. At the scene of a burglary, a latent fingerprint was detected on the doorknob exiting the home. %  where glass fragments were found on the clothing of a person of interest that are thought to have come from a broken window at the house where the burglary occurred.  
Exemplar prints were collected from a person of interest, and Bert the expert was called in to analyze the evidence and provide his opinion.  Alice is interested in deciding between the following two propositions: 

\indent $H_1:$ The latent print recovered from the doorknob came from a finger belonging to the person of interest.\\
\indent $H_2:$ The latent print recovered from the doorknob did not come from a finger belonging to the person of interest.

Prior to receiving Bert's opinion, Alice assesses her prior odds regarding the truth of $H_1$. Upon hearing Bert's opinion, she will apply Bayes' rule to update her uncertainty regarding the truth of $H_1$.  This will require Alice to  evaluate how likely Bert would be to express the given opinion  under $H_1$ and under $H_2$, respectively.

In terms of propositions, these examples reflect the simplest case, each involving exactly two propositions of interest.  This means Alice can update her prior odds using her likelihood ratio for her newly received information, namely the opinion Bert stated.  Alice's $LR$ can be written as 
\begin{align}
LR_A=\myfrac{Pr_A[E_A|H_1,I_A]}{Pr_A[E_A|H_2, I_A]}.
\label{eq. Alice LR}
\end{align}
In this expression, $E_A$ represents Alice's understanding of Bert's opinion, $Pr_A[E_A|H_1,I_A]$ represents how likely Alice feels Bert would be to express that opinion if $H_1$ were true, and $Pr_A[E_A|H_2,I_A]$ represents how likely Alice feels Bert would be to express that opinion if $H_2$ were true.   %If there were more than two propositions of interest, Alice would evaluate Bayes' factors, which depend on the prior probabilities she gives each proposition and the likelihoods of Bert's for the newly received information, to update her uncertainties. 

The steps Alice follows do not depend on what type of opinion Bert provides.  Bert could present his findings  as a categorical conclusion, a numeric likelihood ratio, a perceived level of similarity, the output of a computer algorithm, etc.  In any case, Alice will respond by assessing her likelihood of having received this new information under each proposition of interest to her. %Critically, Bayesian recipients do not view offered opinions as the logical solution to how they should feel about the evidence or simply accept them at face value. Instead, they would consider others' opinions as new information and use their own critical thinking to assess what that new information means to them.  

%\st{Recipients who are previously unfamiliar with the analytic methods used in a particular case may have substantial uncertainties regarding what type of results the methods tend to produce under different scenarios.  For a Bayesian recipient, this uncertainty manifests as ``vague'' or ``non-informative'' prior distributions for the outputs generated by the chosen analytic methods under each of the propositions of interest to them.  The effect of having vague or non-informative priors for the distribution of outcomes under each proposition is that the marginal densities of outcomes themselves are also vague.  This means that the likelihood associated with any given output value will not differ all that much across the considered propositions.  Consequently, the corresponding likelihood ratios or Bayes' factors tend to be close to one and the offered opinion does not substantially affect the Bayesian's posterior probabilities.   If this were really the expectation of recipients, one might ask whether the opinions are worth presenting at all. }

Evaluating likelihoods for each proposition raises another component of uncertainty (in addition to which of $H_1$ and $H_2$ is true).  Namely, what distributions will Alice use to assess these likelihoods?  This choice directly affects the weight Bert's opinion can have on Alice's uncertainty regarding which of $H_1$ or $H_2$ is true.  For Bert's opinion to substantially impact Alice, she must believe the likelihood of Bert expressing this opinion differs greatly depending on which of $H_1$ and $H_2$ is true.   

As a layperson, Alice is unfamiliar with the technical and specialized methods Bert uses as a forensic expert and does not have a strong feel for what type of opinions Bert would express in different scenarios.   Alice reflects her initial skepticism or lack of knowledge about Bert's capabilities by specifying ``vague'' or ``non-informative'' priors.  Such priors allow for a wide range of potential distributions for what opinions Bert would provide under each of the propositions of interest.  Because she does not have a good idea of what opinions to expect under either proposition, Alice's likelihoods for any particular opinion do not differ much between $H_1$ and $H_2$ (i.e., $Pr_A[E_A|H_1,I_A] \approx Pr_A[E_A|H_2, I_A]$). This means Alice will not find Bert very persuasive, regardless of what opinion he provides.

Bert explaining how he arrived at his opinion has little effect on Alice's uncertainties because Alice is not sure how the technical details show which distributions best reflect the opinions Bert would provide in different scenarios.  %Though it may impress them, there is a risk that it will only confuse rather than inform and empower the recipients.  Technical details are only helpful to the extent that they reduce the Bayesian recipients' uncertainty regarding the distribution of outcomes under the various propositions they consider.  If recipients do not comprehend the technical details or their implications, these descriptions may, and should, have little effect on their uncertainty regarding the distributions of outcomes.  
Fortunately, Alice is able to understand descriptions of what opinions Bert has produced in ground-truth-known instances representing Alice's propositions of interest.

Suppose  Bert includes validation data $V$ when presenting his opinion for the case at hand. Alice's new information is now her understanding of the opinion Bert stated and the validation data (i.e., she will consider both $V$ and $E_A$). She will compute her likelihood ratio, $LR_A$ according to Equation~\ref{eq:Alice LR1}
\begin{align}
LR_A=\myfrac{Pr_A[E_A, V|H_1,I_A]}{Pr_A[E_A,V|H_2, I_A]}.
\label{eq:Alice LR1}
\end{align}
Alice believes the validation data is independent of which proposition is true in the case at hand. In this situation, Equation~\ref{eq:Alice LR1} can be rewritten  as
\begin{align}
LR_A=\myfrac{Pr_A[E_A|H_1,V,I_A]}{Pr_A[E_A|H_2,V, I_A]},
\label{eq:Alice LR2}
\end{align}
which reflects treating validation data as accepted and essentially previously known by Alice.  Alice could use either Equation \ref{eq:Alice LR1} or \ref{eq:Alice LR2}  to compute her $LR$.   
These equations highlight the critical role of validation data  and reflect Alice's use of scientific reasoning. She does not, by default, assume that Bert is able to discern which of  $H_1$ and $H_2$ is true well enough to arrive at any particular opinion far more often in one scenario than the other.  Given enough data to demonstrate this ability, however, Alice's revised likelihoods may differ substantially across the propositions she considers (i.e., $Pr_A[E_A|H_1,V,I_A] >> Pr_A[E_A|H_2,V, I_A]$ or $Pr_A[E_A|H_1,V,I_A] << Pr_A[E_A|H_2,V, I_A]$), leading her to give  Bert's opinion substantial weight. 

To limit flow disruptions, some technical details such as precise prior specifications and computational details are provided in Appendix C.  Additionally, to limit the complexity of the illustrations, we assume that, in Alice's judgement, any presented validation data comes from scenarios representative of the current case.\footnote{Assessments of the extent to which a given validation test is representative of a particular case scenario is personal.  Experts could assist stakeholders by explaining what factors are predictive of the distribution of examination outcomes (e.g., expert opinions).}

%transform her vague priors for the distributions of expert opinions into posteriors that are highly differentiated across the propositions in which she is interested.

%As this example will show, validation data can directly reduce this second type of uncertainty, which in turn impacts the effect that Bert's opinion has on Alice's uncertainty regarding which of $H_1$ or $H_2$ is true.  

\subsection{Example with Expert's Conclusion Using a 3-Point Scale}

We first consider a scenario in which Bert provides his opinion as a categorical conclusion by picking one of ``identification'', ``inconclusive'', or ``exclusion''.  %``identified (type-I association)'' (see \citet{totrace}).  
As previously emphasized, Alice does not need Bert to provide a likelihood ratio in order to apply Bayes' Rule.  She simply needs to assess the likelihood of Bert's offered opinion  under $H_1$ and under $H_2$, respectively. Assuming she has no previous exposure to validation data regarding Bert's past responses in scenarios relevant to $H_1$ or $H_2$, Alice has substantial uncertainty regarding how frequently Bert offers various conclusions under these scenarios.  

 For notational simplicity, let \pid, \pinc, and \pex\  denote the expected proportion of comparisons under $H_1$ that Bert would conclude ``identification'', ``inconclusive'', and ``exclusion'', respectively. Similarly, let \qid, \qinc, and \qex\  provide the corresponding expected proportions under $H_2$. After thinking a while, Alice settles on the following assumptions (in addition to the necessary constraints that $\mbox{\pid} + \mbox{\pinc} +\mbox{\pex} = 1$ and $\mbox{\qid} + \mbox{\qinc} +\mbox{\qex} = 1$): 
\begin{itemize}
    \item Identifications among mated comparisons are more common than either exclusions among mated comparisons or identifications among nonmated comparisons (i.e., \pid \ $>$ \pex \  and \pid\  $>$ \qid).
    
    \item Exclusions among nonmated comparisons are more common than either identifications among nonmated comparisons or exclusions among mated comparisons (i.e., \qex\ $>$ \qid\ and \qex\ $>$ \pex).
    
    \item The ratio between the rate of inconclusives among mated and nonmated comparisons, respectively, is smaller than the ratio for identification conclusions and larger than the ratio for exclusion conclusions \bigg(i.e., $\myfrac{\mbox{\pid}}{\mbox{\qid}} > \myfrac{\mbox{\pinc}}{\mbox{\qinc}} > \myfrac{\mbox{\pex}}{\mbox{\qex}}$\bigg).
\end{itemize}
 These constraints correspond to Alice's belief that Bert has some ability to discriminate between $H_1$ and $H_2$ and that his opinions of 'Exclusion', 'Inconclusive', and 'ID' provide increasing levels of support for $H_1$.  Aside from these constraints, Alice is otherwise indifferent in that she feels any combination of \pid, \pinc, \pex, \qid, \qinc, and \qex\ that satisfies the above constraints seems equally fitting. This specifies Alice's prior for the distribution reflecting what opinions Bert would tend to give under $H_1$ and $H_2$ scenarios, respectively. 

Suppose that Bert reports  an identification conclusion for the case at hand.
Even with substantial uncertainty regarding how likely Bert would be to report an ID under either $H_1$ or $H_2$, Alice can still evaluate her likelihood ratio, $LR_A$, for Bert's opinion using what are called marginal likelihoods.  In this instance, the marginal likelihoods are the average values of \pid\ and \qid\ under the prior distribution Alice specified.  As shown in Appendix-C, Alice computes the marginal likelihoods of hearing ``ID'' as 8/15 under $H_1$ and 2/15 under $H_2$, leading to a likelihood ratio of 4 ($=\frac{8/15}{2/15}$).
To give Bert's opinion greater weight, Alice requires information that reduces her uncertainty regarding how often Bert would report ``ID'' under scenarios representing $H_1$ and $H_2$, respectively.  That is, Alice needs performance data regarding Bert's past opinions in ground truth known scenarios she can relate to the current case.   

%\begin{comment}
Suppose Bert provides the following table of validation testing results in addition to his conclusion \footnote{This table summarizes results among the ``value for identification"-rated comparisons in the Noblis black box study for latent print examiners \parencite{ulery2011accuracy}.}

\begin{center}
\begin{tabular}{c|c|c|c}
     Comparison Type & Identification & Inconclusive & Exclusion  \\
     \hline
     Mated& 3663 (61.4\%)& 1856 (31.1\%) & 450 (7.5\%) \\
     Non-mated& 6 (0.147\%) & 455 (11.1\%) & 3622 (88.7\%)
\end{tabular}
\end{center}

As shown in Appendix-C, Alice's revised $LR$ after encountering the validation test results is 358, which is a substantial increase over her initial $LR$ of 4 based on Bert's opinion alone before encountering the validation test results. 

\begin{figure}[H]
    \centering
    \includegraphics[width=\linewidth]{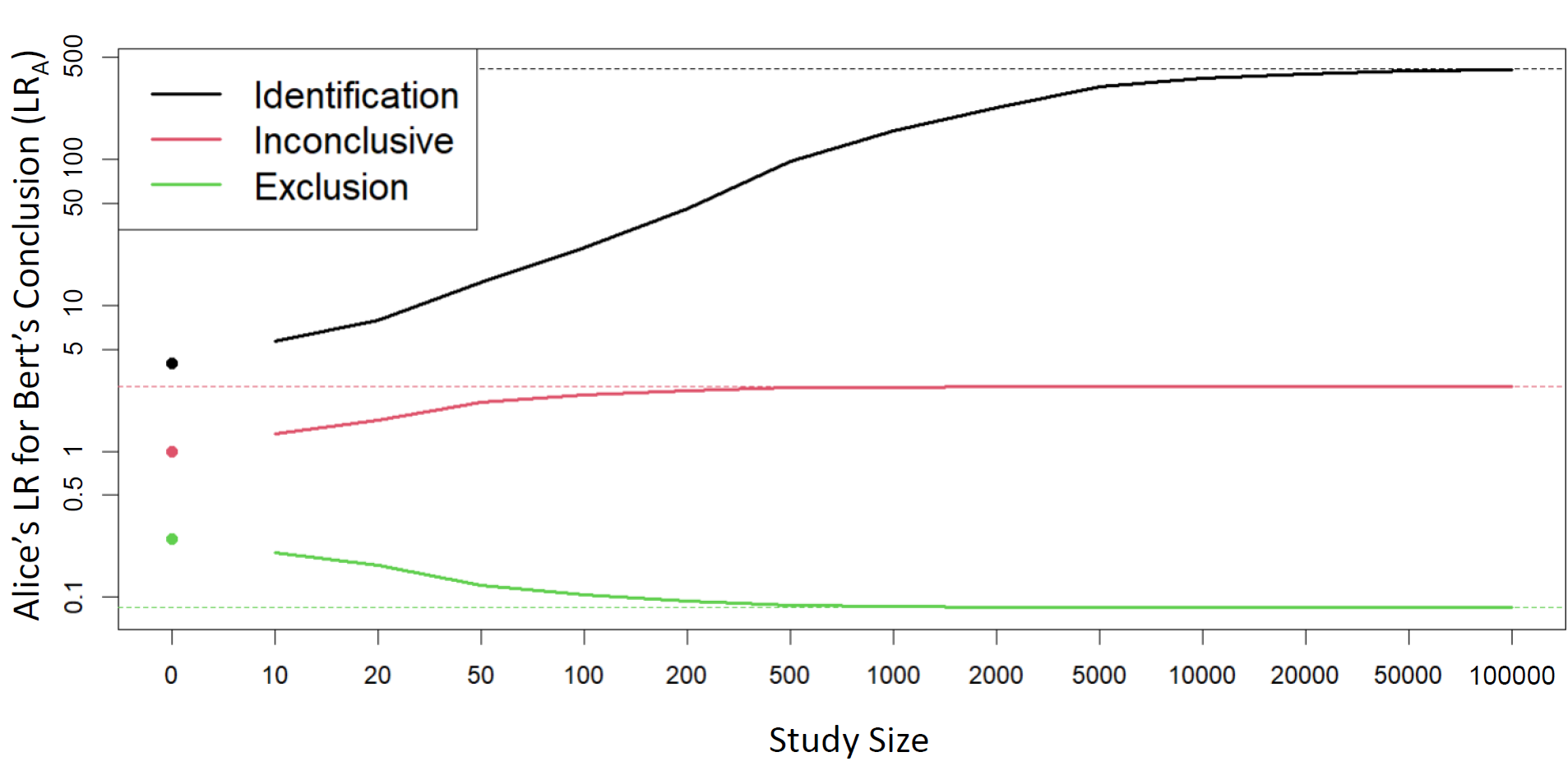}
    \caption{Effect of validation sample size on Alice's $LR$ for Bert's conclusion.  Dashed horizontal lines indicate the ratio of observed conclusion rates in the provided validation data.} 
    \label{fig:CategoricalLR}
\end{figure}

The effect of the validation data on Alice's $LR$ depends on how many observations are included in the validation testing.  Figure \ref{fig:CategoricalLR} shows Alice's $LR$ for each conclusion level as a function of study size, using the same proportion of mated (59.4\%) and non-mated (40.6\%) comparisons and keeping the conclusion rates within each comparison type the same as in the original study results.  

As one would expect, Alice's $LR$ for each conclusion type converges to the ratio of the observed conclusion rates as the study size increases (e.g., for identification, the observed ratio was 61.4\% / 0.147\% $\approx$ 418). These asymptotic curves illustrate both the continuously increasing benefit and eventual diminishing return of additional validation testing.  In this example, Alice's $LR$s for inconclusive and exclusion opinions change very little after a study with about 500 samples, while her $LR$ for an identification opinion continues to increase even after a study with 5000 samples.   Ultimately, choosing how much validation data to collect is a cost-benefit decision.  There is no sample size where a method suddenly shifts from unvalidated to validated.  In general, it will take more data to support more extreme $LR$s (i.e., $log(LR)$s further from 0).

\subsection{Example with Expert's Value of \texorpdfstring{$LR$}{TEXT}}
     
Instead of using a categorical conclusion, this time Bert summarizes his findings using a likelihood ratio, $LR_B$, for the same pair of propositions Alice considers.  We use the notation $LR_B$ to indicate that it is the strength of evidence according to Bert, where
$$
LR_B = \myfrac{Pr_B[E_B | H_1, I_B]}{Pr_B[E_B | H_2, I_B]}.
$$
The notation $Pr_B$ indicates Bert's probability assessments.    

Alice's process closely follows the first example. The only substantial change will be that Alice uses different distributions to represent her uncertainty in what opinions Bert would provide under $H_1$ and $H_2$, respectively.  This change is necessary because $LR$ opinions are continuous (taking any value from zero to infinity) whereas categorical opinions are discrete. 

Suppose Bert explains the propositions and evidence he considered and informs Alice that his weight of evidence, $log_{10} LR_B$, is $r$ (i.e., $LR_B=10^r$). % and that Alice understands this information (i.e., $E_A$ is the fact that Bert said his weight of evidence for the given propositions is $r$). 
Upon hearing Bert's $LR$, Alice is interested in assessing her updated belief regarding the truth of $H_1$, which can be written as $Pr_A[H_1|log_{10} LR_B=r, I_A]$. This requires Alice to assess the likelihoods that Bert would say $log_{10} LR_B=r$ under $H_1$ and  $H_2$, respectively.  

Alice has never seen any data regarding what values Bert (or any other expert for that matter) has produced in the past when using Bert's chosen method for evaluating a $LR$.  Thus, Alice is unsure what distributions reflect the weights of evidence Bert would provide when analyzing two impressions from the same finger %glass fragments from a common source 
(corresponding to $H_1$) or when analyzing two impressions  from different sources (corresponding to $H_2$).  Alice describes her uncertainty regarding the distribution of $log_{10} LR_B$ under each proposition using blends of many normal distributions with different means and variances. This can be considered as Alice's prior for the potential $log_{10} LR_B$ distributions under $H_1$ and $H_2$. (Exact details of Alice's prior are provided in Appendix-C.)  

The marginal distribution Alice needs in order to rationally assess how likely she feels Bert would be to say $log_{10}LR_B=r$ under $H_1$ is the weighted average of all the normal distributions in Alice's prior for $H_1$, where the weights for each normal distribution come from her prior distribution. A marginal distribution for $log_{10} LR_B$ under $H_2$ can be evaluated in the same way.

\begin{figure}[H]
    \centering
    \includegraphics[width=.5\linewidth]{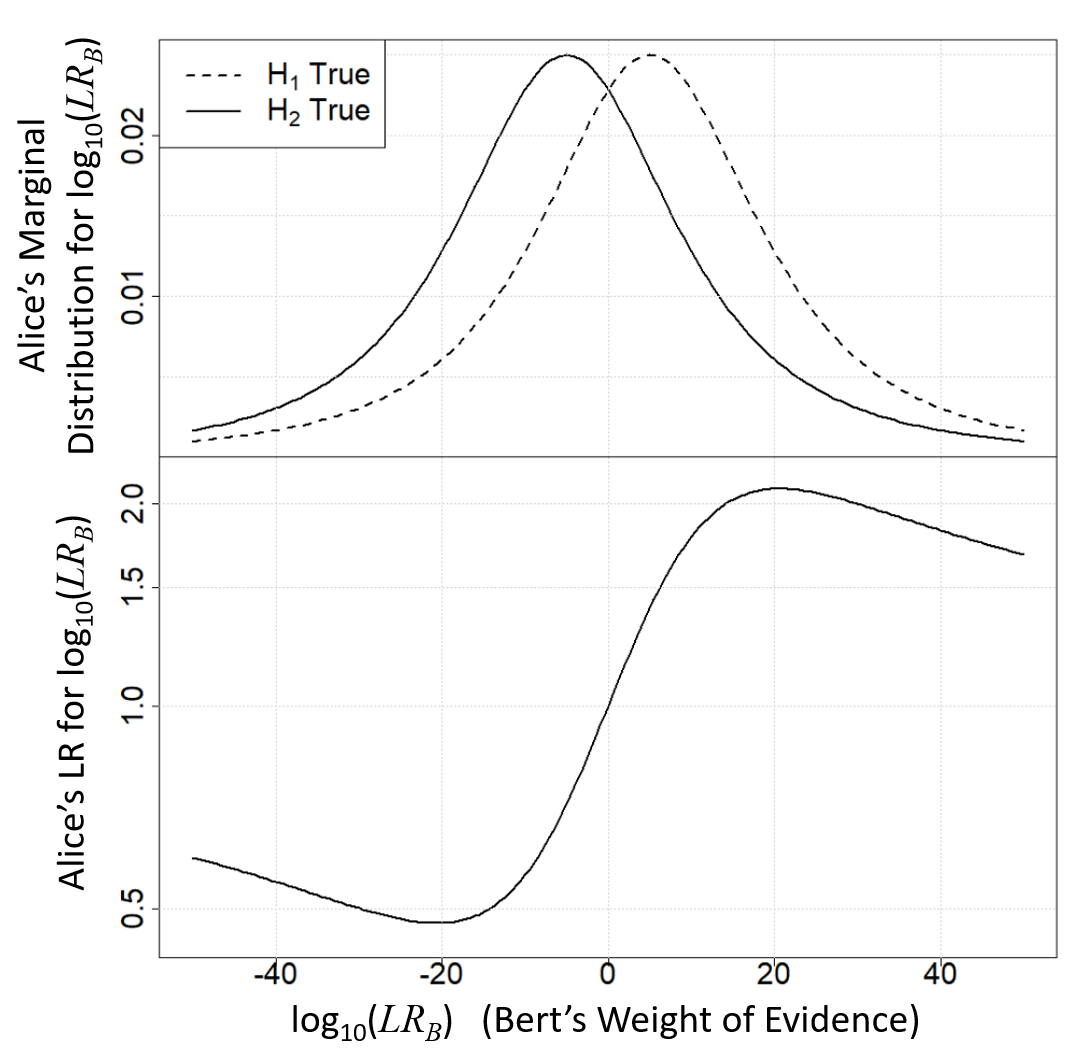}
    \caption{Top: Alice's initial marginal distributions for $log_{10}(LR_B)$ under $H_1$ (dashed curves) and under $H_2$ (solid curves). Bottom: Alice's $LR$ ($LR_A$) as a function of the weight of evidence Bert reports ($log_{10}LR_B$).}  
    \label{LR_prior}
\end{figure}

The marginal distributions for $log_{10} LR_B$ under each proposition according to Alice's priors are shown in the top panel of Figure \ref{LR_prior}. Alice can evaluate her $LR$ for the observation that Bert said $log_{10} LR_B=r$ by dividing the density at $log_{10} LR_B=r$ in the marginal distribution for $H_1$ by the density at $log_{10} LR_B=r$ in the marginal distribution for $H_2$.  The resulting ratios are plotted in the bottom panel of Figure \ref{LR_prior}.

Notice that for the priors Alice has chosen, even extreme values of $LR_B$ will have little influence on her uncertainty regarding the truth of $H_1$.\footnote{The exhibited behavior where $LR_A$ shrinks towards 1 for extreme values of $LR_B$ is a consequence of the blend Alice chose to represent her uncertainties for the distribution of $LR_B$ under $H_1$ and $H_2$.    Though somewhat counter-intuitive, this effect is irrelevant to the point of this example, which was chosen for its computational simplicity.} For instance, if Bert states that his $LR$ is a billion (i.e., $log_{10}LR_B=9)$, Alice's $LR$ would only be about 2.  For Bert's opinion to have greater impact, Alice must also receive information that reduces her uncertainty regarding what values of $LR_B$ Bert would provide under $H_1$ and under $H_2$.  

Suppose Bert, in addition to providing $LR_B$ for the case at hand, also provides results from validation testing where he was asked to evaluate $LR$s in reference scenarios where a third party knew whether or not the two fingerprint impressions being evaluated were from the same source.  Suppose among the provided results there are $n_1$ tests that Alice  views as having come from the same distribution as $LR_B$ would have if $H_1$ were true.  Taking these validation test results into consideration reduces Alice's uncertainty regarding what  $log_{10}LR_B$ values Bert tends to report under $H_1$. 

\begin{figure}[H]
    \centering
    \includegraphics[width=.6\linewidth]{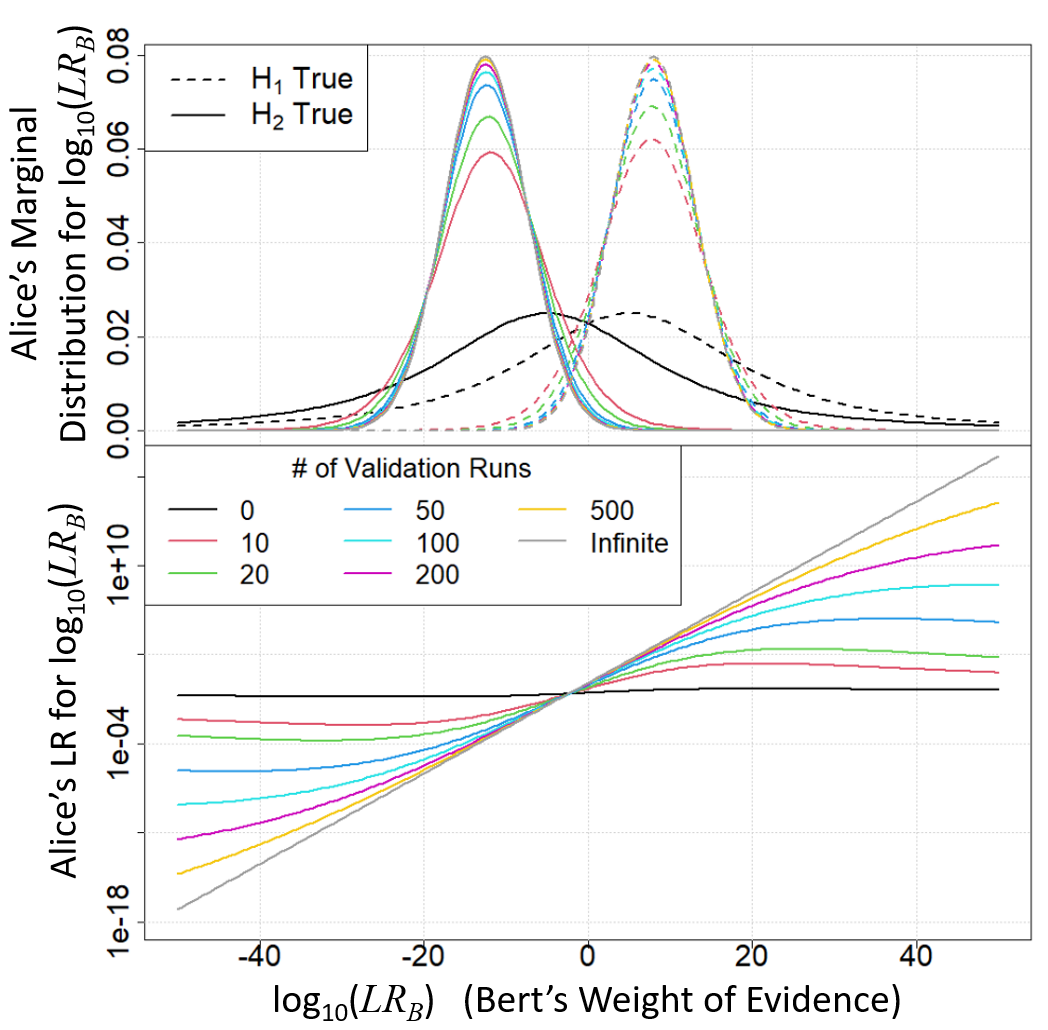}
    \caption{Top: Alice's marginal distributions for $log_{10}(LR_B)$ under $H_1$ (dashed curves) and under $H_2$ (solid curves) for varying numbers of validation tests. Bottom: Value of Alice's $LR$ ($LR_A$) as a function of $log_{10}(LR_B)$ after Bert discloses results from varying numbers of validation tests.}
    \label{fig:LR_post}
\end{figure}

This effect is shown in the top panel of Figure \ref{fig:LR_post} (dashed curves) for various numbers of validation tests, with each considered sample having a mean of 8 and a variance of 25.  As one would expect, the larger the collection of validation samples provided to Alice, the more strongly her marginal distribution for $log_{10}LR_B$ under $H_1$ is pulled toward the observed attributes of the validation data (i.e., a mean of 8 and variance of 25). 
The solid curves in the top panel of Figure~\ref{fig:LR_post} reflect a parallel exercise that considers the effect of Bert providing Alice with $LR$s from validation tests conducted in scenarios that she views as having come from the same distribution as $LR_B$ would have if $H_2$ were true.  For these computations, the validation samples for $H_2$ were considered as having a mean of $-12.5$ and a variance of $25$.

The bottom panel of Figure~\ref{fig:LR_post} depicts the value of Alice's $LR$ ($LR_A$) as a function of Bert's stated evidential weight ($log_{10}LR_B$) following disclosure of various numbers of validation $LR$s.  As expected, providing additional validation test results increases the potential effect that $LR_B$ can have on Alice's uncertainty regarding the truth of $H_1$.

\subsection{Example with Expert's \texorpdfstring{$LR$}{TEXT} Range}

In this scenario, Bert provides his opinion using an uncertainty interval for a $LR$ rather than a single value. For instance, rather than providing a likelihood ratio of a billion, Bert provides a range for a likelihood ratio of 100 million to 10 billion.  Consideration of this scenario is motivated by the following excerpt from the article ``The $LR$ Does Not Exist''  \citep{berger2016lr}: ``... we think there is no rational way to use such an interval, if presented with one. We invite those that propose to report an $LR$ with an interval to demonstrate how one should update one's prior odds into posterior odds, based on that interval, for the purpose of decision making."

%Charles Brenner put forward a similar challenge in 2000 on\\ \centerline{http://dna-view.com/noconfid.htm\#challenge}\\ and reports that no scientist has come forward to explain how to rationally use an interval (for profile frequency) in decision making.

Alice handles this scenario by applying Bayes' rule as in the previous examples.  $LR$ intervals only become problematic if one envisions a deferential recipient who would try directly inserting Bert's range  into Bayes' rule in place of their own $LR$.  This would produce a range of posterior odds rather than a single value, which could result in situations where the interval endpoints would lead a recipient to opposite decisions. The problem in this scenario is not that Bert has provided an interval, but that a recipient is attempting to use the ``deferential-Bayes" equation, which is a faulty application of Bayesian reasoning.  %We now illustrate how someone could actually apply Bayesian reasoning in response to learning an expert's $LR$ interval.

At a high level, little has changed from the previous example where Bert provided $LR_B$ as a single value.  Alice would not feel a constraint that her interpretation of the evidence must fall within Bert's range.  Instead, she would consider Bert's provided range as new information to be processed using Bayes' rule in accordance with her own personal uncertainties.  In particular, Alice would consider the likelihood of Bert having provided the given range under each  proposition of interest.  Similar to the first examples, Alice has little idea regarding what distribution would represent the ranges Bert would report under the different propositions of interest without having access to relevant performance data.  Substantial uncertainty of this type would limit the potential influence that Bert's provided range has on the Alice's uncertainty regarding which of the considered propositions is true.  Bert could increase the potential utility of his provided interval to Alice by also providing performance data to illustrate what ranges have been obtained in past instances for scenarios representing similar propositions of interest to those in the case at hand.  Appendix-C goes through the computational exercise of illustrating the effect of such performance data and largely proceeds in the same manner as the previous examples except that Alice now needs to formulate prior distributions over the space of bivariate distributions for each proposition of interest.  This increases the computational complexity compared to dealing with a scalar or categorical opinion as in the previous examples. 

This example is intended to illustrate that, in theory, recipients can apply Bayesian reasoning to update their uncertainties in response to any type of new information, including intervals.  It should not be interpreted as a recommendation that experts communicate using intervals, which would depend on the expert's intention for how a recipient will use their opinion.  If the goal is for experts to help recipients establish their own weight for an expert's opinion by providing performance data, there are clear benefits to using the simplest forms of interpretations (e.g., a single number or category). This is consistent with the perspective that a primary role for experts is to transform complex physical evidence into information that is as simple as possible for the recipients so that they have an easier time deciding what that information means to them. The added difficulty in processing more complex information (such as uncertainty intervals or distributions in place of point estimates) should be weighed against the expected influence the additional information would have on recipients' uncertainties.

On the other hand, if an expert is not intending to provide performance information that would empower recipients to make their own assessments of what the expert's opinion means, then it seems advisable to acknowledge that the expressed opinion is one from a range of opinions and to describe the extent to which that range of opinions has been explored (e.g., by fitting multiple models or asking multiple experts).

\subsection{Example with Two Experts Providing \texorpdfstring{$LR$s}{TEXT} }

In previous examples Alice responded to information presented by a single expert, Bert, such as in a report issued before a trial occurs. When a case proceeds to trial, an expert's opinion is often accompanied by competing perspectives. For instance, suppose Bert is nominated by the prosecution.  The defense could nominate their own expert, Carla, to discuss the same evidence Bert considered, or even offer opinions related to the same two propositions $H_1$ and $H_2$. Suppose Carla  offers her opinion also in the form of her personal $LR$, which we write as $LR_{C}$. 

When Bert and Carla provide $LR_{B}$ and $LR_{C}$, respectively, for the same evidence and the same considered propositions, it is impossible for Alice to accept both values as her own unless $LR_{B}$ and $LR_{C}$ are equal. The important implication is that the deferential-Bayes equation (Equation~\ref{eq:BayesHybrid}), which some papers continue to promote (e.g.,  \textcite{biedermann2018analysing}, \textcite{aitken2018roles} and \textcite{aitken2018commentary}), cannot accommodate the common scenario of experts providing differing opinions. Additionally,  instances where Alice receives new information or considerations revealed by cross-examination also fall outside what can be represented through an application of Equation~(\ref{eq:BayesHybrid}).  

In the appropriate application of Bayes' rule, all components of uncertainty belong entirely to the recipient (Equation~\ref{eq:BayesRational}), who could consider the entirety of the newly available information and form their own likelihood under each of the propositions of interest to them.  In the instance where Bert provides $LR_{B}$ and Carla provides $LR_{C}$ and Alice is interested in propositions $H_1$ and $H_2$, Alice would need to assess her probabilities $Pr_A[H_1 | LR_{B}, LR_{C}, I_A]$ and $Pr_A[H_2 | LR_{B}, LR_{C}, I_A]$. This can be done by assessing her own $LR$ as 
\begin{align}
    LR_A = \myfrac{Pr_A[LR_{B}, LR_{C} | H_1, I_A]}{Pr_A[LR_{B}, LR_{C} | H_2, I_A]}.
\end{align}
and using Bayes Rule to update her prior odds. There is no mathematical reason for  $LR_A$ to equal either $LR_{B}$ or $LR_{C}$.  We provide computational details for Alice's assessment in this scenario in Appendix-C.

%Ultimately, there is a trade off between the added complexity of interpreting additional information and the potential benefit of doing so.

%On the other hand, if performance data is not provided to recipients, the extensiveness of the uncertainty evaluation used to produce the provided interval could directly impact the potential influence of the provided interval.

%If an expert has conducted a minimal uncertainty analysis (e.g., considering only the variability that may occur if the same modeling approach were applied to a new sample of data the same size as what is currently available), one might expect a Bayesian's prior distributions for bivariate distributions to exhibit a similar level of discrimination across the propositions of interest as prior distributions for the distributions of a single provided value.

%As the extent of an expert's uncertainty characterization increases, the potential impact of intervals with lower bounds far above 1 or upper bounds far below 1 may increase.

\section{Discussion}

Some authors have said that Bayes' equation shows an examiner's role is to provide a likelihood ratio (e.g., \cite{robertson2016interpreting, aitken2008fundamentals, aitken2018roles}).  We have previously argued against this perspective \parencite{lund2017likelihood}.  In the current paper, we  distinguish deferential recipients from logical and scientific recipients and reiterate that, as shown in Morris' framework \citep{morris1971bayesian,morris1974decision,morris1977combining}, Bayesian reasoning does not support deferential recipients.  

While we approached this subject  from the perspective of Bayesian reasoning, the question of whether experts should educate recipients or recipients should defer to experts has been discussed from a legal perspective for decades. See, for instance,  \citet{epstein1992judicial,allen1992common}.  Normative applications of Bayesian reasoning clearly favor experts educating recipients.  One distinction that comes from the Bayesian perspective is that recipients need not necessarily be educated on the scientific foundations of a discipline.  Ultimately, recipients must decide what weight to give an opinion or result. 
While understanding technical details for how an expert's method works could be helpful, it is more direct, and scientifically sound, to consider the demonstrated performance of the methods used by the expert.

For instance, a recipient may not understand the science behind laser-ablation-inductively-coupled plasma-mass spectrometry (LA-ICP-MS) well enough to have a good sense of how often experts using the methodology might mistake glass shards from one manufacturer as having come from a different manufacturer.  However, even a recipient who does not know what LA-ICP-MS stands for could understand a statement that experts using LA-ICP-MS made no such mistakes in 100 blind tests in which they received glass shards from one manufacturer and were asked to compare them to glass from another manufacturer.

%A general framework describing how a logical and scientific recipient can interpret expert opinions is provided in \citet{morris1971bayesian,morris1974decision,morris1977combining}. 

%We have further clarified that, when pursuing logical and scientific communication, an expert should provide validation results so recipients can better assess the weight of the expert's findings. 
%The examples we provided emphasize the importance of experts  providing information to recipients that will help them assess the weight of the expert's findings. 
Through several examples, we have shown the important role of validation data in the normative process for assessing what weight to give an expert's opinion.  The computations provided in the appendix are intended for theoretical rigor and clarity, but we acknowledge recipients are unlikely to perform such explicit calculations.  This does not take away from the general argument. In a system founded on logic and science, recipients generally should not be expected to give an expert's opinion much weight without compelling demonstrations of expert performance.  How much weight they give would depend on how strong the validation data is.  In that sense, validation data is just as important to logical and scientific discourse as the expert opinion itself.  This reinforces interest in the question of how best to present validation data to recipients.  

In theory, providing complete details of every validation test and result would allow recipients to extract all the information relevant to their assessments.  Practically, it is not so simple.  Data quickly becomes overwhelming and some form of summary may be  necessary.  Summarizing data, however, is also a subjective exercise, and perhaps should not be left to any one person, even an expert.   

One approach would be for experts to aggregate all available performance data for their chosen methods into one or more spreadsheets and to make these available to interested parties (e.g., as an attachment or URL)  as part of the report in which they provide their opinion.  In adversarial situations, representatives of interested parties could highlight pertinent aspects of the data the expert has provided to help recipients assess what weight they will give the examiner's opinion.  Making the data publicly available would also facilitate independent reviews of method performance and potentially lead to improved methods.

Note that by performance data, we are not meaning summaries in the form of error rates.  Error rates are  summaries of performance data, but their suitability as performance measures for experts using opinion scales with more than two levels has been rightly questioned \parencite{weller2020commentary}.  Additionally, computing error rates requires averaging across various use cases, which may or may not be appropriate in eyes of the recipient.  Rather, by performance data, we mean a table of results where each row corresponds to one application of a method during testing and columns are used to specify the circumstances under which that application occurred and what results (or opinions) were obtained. Some black box studies have already included such tables of results (e.g., \cite{hicklin2022accuracy,eldridge2021testing}).   Where possible, it is also helpful to provide what was compared (such as impression images) so that stakeholders can consider factors related to quantity and quality of information not represented in the contents of the spreadsheet.

The term ``validated'' is often used when discussing method performance.  This term conveys whether a person or organization has reviewed, and feels comfortable with, the performance of a given method.  This paper has emphasized the importance of recipients considering  demonstrated method performance for themselves rather than adopting someone else's perspective as their own.  Performance assessments and thresholds are subjective, and knowledge regarding method performance is not binary.  As shown in first two examples of Section~3, each bit of testing provides a little more insight into the performance of the studied method, but at no point is the knowledge complete. There is always an ongoing cost-benefit trade-off for additional testing. Demonstrating higher levels of performance requires more testing. There is no threshold of data beyond which logical recipients should accept  an expert's opinion as their own.  Logical and scientific recipients must assess their personal uncertainties after considering how much testing has been done.  These basic principles are lost when focusing on whether or not a method has been validated rather than focusing on what information is available regarding the method's performance.

To be clear, we are not decrying methods that rely on expert opinions or judgement. Experts play a critical role in identifying, collecting, and analyzing evidence.  Without the skills of forensic experts, other members of the judicial system would have to attempt to deal with complex and chaotic crime scenes and highly technical applications of chemistry, physics, and biology that require years of training and experience to perform correctly. %They would have to guess when sifting out subtle discriminating features and comparing complex patterns awash in a sea of variability and interference.  No one is suggesting that other members of the judicial system are capable of performing the analysis of evidence themselves. 
Recipients, logical or otherwise, trust and depend on trained experts to transform raw information in the form of physical or digital evidence into a much simpler scale others can understand and to do so in the most effective way among any methods discovered to date.  %Oftentimes, the most effective methods currently available for discriminating between $H_1$ and $H_2$ (e.g., whether or not two footwear impressions were made by the same shoe) we have to date  are the examiner's experience-based opinions themselves.  
This paper considers how evidence communication can best support an examiner's opinion or result (e.g., algorithm output) in a logical and scientifically sound manner.  To that end, we are advocating for factual reporting and testimony about method performance, regardless of the extent to which that method depends on expert opinions and judgments.  Ultimately, we hope these perspectives help shape continuing conversations regarding how experts should communicate with other members of the judicial system. %We suggest that assessments of different approaches to expert communication center around whether the information provided is accurate, and whether recipients find the information to be understandable and helpful.  

\vskip 0.2in
\subsubsection*{Disclaimers:} This work was created by employees of the National Institute of Standards and Technology (NIST), an agency of the Federal Government. Pursuant to title 17 United States Code Section 105, works of NIST employees are not subject to copyright protection in the United States. 

These opinions, recommendations, findings, and conclusions do not necessarily reflect the views or policies of NIST or the United States Government.

\vskip 0.2in
\subsubsection*{Acknowledgements:} We thank John Butler, Jan Hannig, Martin Herman, and Yooyoung Lee for their valuable comments. %on an initial draft of this paper.  
We especially thank Will Guthrie for support and valuable feedback while reviewing multiple drafts of this paper.

\pagebreak
\setcounter{biburlucpenalty}{8000}
\printbibliography
\pagebreak
\section*{Appendix-A}
\label{sec:Appendix-A}
\vskip -0.25in
\noindent{\bf \Large  Probability is Personal}

Probability theory is extremely useful for an individual who has to make decisions in the presence of uncertainty. Unfortunately, one person's probability does not transfer to another person because how uncertain one person feels regarding the truth of a proposition is not necessarily how uncertain someone else would, or should, feel about it. The feeling of uncertainty is highly subject specific. Different individuals can, and generally do, have different degrees of belief about the same event or proposition, even if they have the same background data or information available. This is because {\em data, by themselves, do not produce probabilities.} 

%\noindent{\bf Probabilities are personal. } 
It has been recognized by many of the founders of modern probability theory that probabilities are personal (\textit{e.g.}, \cite{lindley2013understanding, kadane2020principles, de2017theory}). According to them, a probability is a quantitative expression of the degree of belief in the truth of a statement (proposition, hypothesis, event) that an individual has based on their current knowledge and other beliefs. 
Kadane says, in the very first chapter of his book titled ``Principles of Uncertainty" the following: 
\begin{quote}
    ``Before we begin, I emphasize that the answers you give to the questions I ask you about your uncertainty are yours alone, and need not be the same as what someone else would say, even someone with the same information as you have, and facing the same decisions."
\end{quote}
The only requirement for a logical and mathematical treatment of such personal probabilities is that the collection of probabilities assigned by an individual to a set of related propositions obey the basic laws of probability. This property is often referred to as {\it coherence}. 

The basic laws of probability themselves can be derived by adopting the ``avoid sure loss'' principle \parencite{kadane2020principles}, that is, one will not make a bet that is known, with certainty, would result in a loss to the person making the bet. Using this argument (sometimes also referred to as a ``Dutch book argument'') the fundamental laws of probability can be derived. As Kadane says (words in italics added by us),
\begin{quote}
Avoiding being a sure loser requires that your prices {\em (probabilities)} adhere to the following equations:
\begin{description}
\item (1.1) $Pr\{A\} \geq 0$ for all events $A$
\item (1.2) $Pr\{S\} = 1$, where $S$ is the sure event
\item (1.3) If $A$ and $B$ are disjoint events, then $Pr\{A \cup B\} = Pr\{A\} + Pr\{B\}$.
\end{description}
If your prices [(i.e., probabilities)] satisfy these equations, then they are coherent.
\end{quote}
Kadane goes on to say,
\begin{quote}
Coherence is a minimal set of requirements on probabilistic opinions. The most extraordinary nonsense can be expressed coherently, such as that the moon is made of green cheese, or that the world will end tomorrow (or ended yesterday). All that coherence does is to ensure a certain kind of consistency among opinions. Thus an author using probabilities to express uncertainty must accept the burden of explaining to potential readers the considerations and reasons leading to the particular choices made. The extent to which the author’s conclusions are heeded is likely to depend on the persuasiveness of these arguments, and on the robustness of the conclusions to departures from the assumptions made.
\end{quote}
In particular, being coherent does not imply being true.

A compelling illustration of the fact that probabilities, and hence likelihood ratios, are personal is offered in the book \parencite{kadane2011probabilistic}.  There the authors use the case of ``... a shoemaker named Nicola Sacco and a fish peddler named Bartolomeo Vanzetti who were charged with first-degree murder in the slaying of a payroll guard during an episode of armed robbery that took place in South Braintree, Massachusetts, on April 15, 1920, to illustrate the multiplicity of issues that arise when considering a complex collection of evidential material and attempt to derive probabilistic conclusions using a chain of plausible arguments. ... " In particular they illustrate (see Chapter 6 of their book) to what extent likelihood ratio assessments made by the authors, and another individual very familiar with the details of the Sacco and Vanzetti case, differ from one another. 

\begin{comment}
{\cred Walk through definition of probability to set table for argument that ``Bayesian theory requires each decision maker provides their own probabilities''. Not just a number between 0 and 1.  Mention the book by Kadane and two others proceeding through the Bayesian analysis of the Sacco(?) case.}
\end{comment}

In reality, different people may assign different values to probabilities for a variety of  reasons they consider to be valid. This is not an issue when each individual is making probability assessments for use in their own decision making, but it is an absolutely critical issue if attempting to tell someone else what their uncertainty should be. Most types of evidence are complex structures with many attributes, each of which could be considered to varying degrees or ignored altogether by different individuals.  A concrete example is provided by the different approaches that are still being practised in DNA mixture interpretation: the binary model, the semi-continuous model, and the continuous model, to name a few, where some models use only a part of the information used by other models. That is, even $E$ (i.e., evidence) by itself is rather ambiguous.  

Further, terms placed to the right of the vertical bar, such as $I$ in $Pr[E|I]$, represent information considered as indisputable fact by the individual forming the probability;   however, individuals may disagree as to what constitutes fact. Just because one person, even an expert, treats something as a proven fact, does not mean all decision makers must.  A decision maker might agree with some portions of what an expert treats as factual and question, or outright reject, other portions. It makes good sense to consider probabilistic models that accommodate these real world situations. Articles, such as  \citet{aitken2018roles}, that ignore these very real considerations by embracing the practice of having decision maker $A$ use $LR_B$ in place of their own $LR$ (i.e., $LR_A$).

For the benefit of readers who have not been previously exposed to the fact that probabilities are personal, we provide an illustrative example below.

\noindent{\bf Example.} Suppose a coin was tossed by a mechanical device eight times and the results were $HHHHHTTT$, in this order. Let us now consider the question ``What is the probability that the result of the ninth toss would be heads ($H$)?"  We illustrate that answers to this deceptively simple question are personal by considering the responses of three different hypothetical individuals, say A, B, and C.  

\begin{figure}[H]
    \centering
    \includegraphics[width=0.9\textwidth]{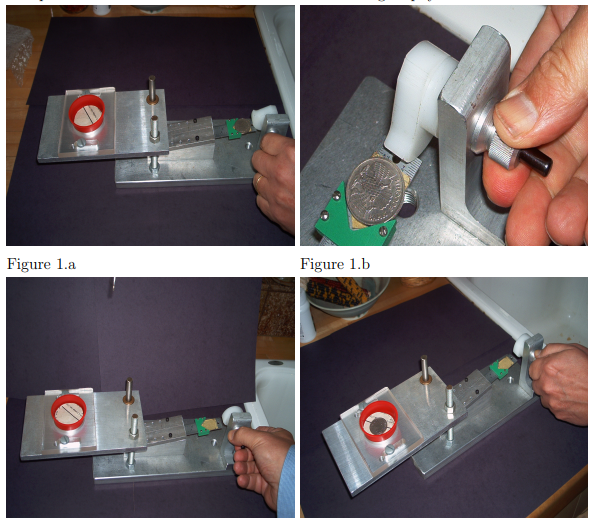}
    \caption{Mechanical coin tossing device used by J. B. Keller. [Keller, 1986] The probability of heads, American Mathematical Monthly, 93:191-197.}
    \label{fig:CoinTossingMachine}
    %Persi Diaconis Paper (uploaded to this folder)
\end{figure}

Individual $A$ believes from the outset that the coin tossing mechanism will be fair and the tosses will be independent and assesses the probability of the ninth toss being heads to be 1/2, regardless of what was observed among the first eight tosses. This corresponds to using a degenerate prior of 1/2 for the probability of heads on any given coin toss. %She may even carry out a diagnostic statistical test to assess whether the observations are inconsistent with her assumed model and note that there is no compelling evidence against her model. 

Individual $B$ feels uncertain about the behavior of the coin-flipping apparatus and represents his uncertainty using a uniform distribution for $p = Pr(Heads)$.  He further assumes that, if  $p$ were known, then the outcomes of individual tosses follow an independent and identically distributed Bernoulli model with probability $p$ of obtaining `heads' in each toss. In particular, the number of heads observed in $n$ flips would follow a binomial distribution with parameters $n$ and $p$. After observing that 5 out of 8 flips resulted in heads, B's uncertainty regarding $p$ follows a beta distribution with parameters $\alpha=6$ and $\beta=4$.  The expected value of this distribution is 0.6, and $B$ assigns a probability of 0.6 to the event that the ninth toss will be heads. 

Individual $C$ views the tossing device and wonders if it might be prone to ``drifting'' such that the forces applied for tossing the coin keep changing gradually from one toss to the next.  To account for this possibility, $C$ does not assume the tosses will all be independent, but instead chooses to represent the probability of heads on a given toss as being dependent on the outcome of the previous toss.  More specifically, $C$ conceptualizes the flipping system using two separate probabilities,  {$Pr($Next flip heads$|$previous flip was heads$)=p$} and {$Pr($Next flip heads$|$previous flip was tails$)=q$}.  Furthermore, $C$ represents her uncertainty about $p$ and $q$ using (mutually independent) uniform distributions.  Among the last seven flips of the observed sequence $HHHHHTTT$ there are clearly four heads and one tails among the five flips that immediately follow an observed heads, and two flips that immediately follow an observed tails, both of which are tails. There is some ambiguity regarding what to do with the first flip, which was heads, because $C$ does not know the outcome of the flip that occurred before it, which we denote $Y_0$. $C$ reflects this uncertainty by assuming $Y_0$ was as likely to have been heads as it was tails. After applying Bayes' rule, the updated uncertainty regarding $p$ is the average of two beta distributions, one with parameters $\alpha=6$ and $\beta=2$ (reflecting the instance where $Y_0$ was heads and the first heads in the observed sequence is included in the total) and the other with parameters $\alpha=5$ and $\beta=2$ (reflecting the instance where $Y_0$ was tails and the first heads in the observed sequence is not included in the total).  Similarly, the updated uncertainty regarding $q$ is the average of two beta distributions, one with parameters $\alpha=1$ and $\beta=3$ (reflecting the instance where $Y_0$ was heads and the first heads in the observed sequence is not included in the total) and the other with parameters $\alpha=2$ and $\beta=3$ (reflecting the instance where $Y_0$ was tails and the first heads in the observed sequence is included in the total).  Because the last observed flip in the observed sequence of flips was tails, C's probability that the ninth toss will be heads is given by the expected value of $q$, which is 0.325. 

Each of the three individuals above has applied Bayes' rule correctly, and therefore each of the three individuals can claim to be logical and coherent. Yet the perceived probability of heads in a ninth flip differs substantially across the three individuals and none of them can be labeled as incorrect. 
Even though all individuals have the same knowledge of the empirical data (the results of the first eight flips), they arrive at different personal probabilities because their initial beliefs were different.  None of their respective mental stories regarding how the mechanical device might behave is inherently more truthful than any other.  Correspondingly, none of their chosen priors are more appropriate than any other.

This example is intended to illustrate the basic fact that, even in simple scenarios, different individuals can follow Bayesian reasoning and arrive at different probabilities for the same propositions given the same data.  The more complex a statistical model becomes, the more opportunities there are for modeling choices to substantially affect the outcomes of model.

The fact that probabilities are personal has a profound implication when it comes to judging the value of evidence or expert opinion, namely, such judgments are also personal. 
In a criminal trial, fact finders may have certain initial beliefs regarding claims made by the parties involved. 
Their beliefs can change based on any new information (factual or opinionative) presented. The extent to which they modify their beliefs as a result of the new information is a personal judgment and will often vary from one individual to another. There is no normative guidance for how much influence a given piece of information should have, except for a notion that the reasoning used by the person making the judgment should be logical and self-consistent. Thus, there is no single correct weight or strength associated with any given piece of information. This  sentiment is also expressed in  \citet{berger2016lr}. 

In general, the subjectivity of probabilistic interpretation has many sources (e.g., confidence in the motives and skills of the persons collecting and processing the evidence in this case, the representativeness of reference evaluations used to inform distributional assumptions for the given case, the actual assumed distributions, etc). 
This subjectivity of probabilistic interpretation should influence our choice in evidence communication strategies.  If we know that interpretations vary across individuals and models, why would we choose to emphasize the interpretation of one expert or one model, especially without thoroughly attempting to understand the level of variability among experts or models in a given case?  In the example above, we suggest that the informational value is entirely contained in the data $HHHHHTTT$ and any background information that might be available regarding the coin tossing device. Hearing the personal interpretation from one person (e.g., who thinks the probability that the next flip results in heads is 0.5 (or 0.6, or 0.325)) does not add any scientifically defensible value. In fact, focusing on a single probabilistic interpretation can be misleading since it does not convey to the recipient that there are many other plausible and equally justifiable assessments. The recipient is left with inadequate information to judge the reliability of the given opinion and, in many cases, may not even be aware of this fact. 
 
\section*{Appendix-B}
Here we list some examples from the forensics literature where authors explicitly promote use of the ``deferential-Bayes" equation. 
 
\begin{itemize}

    \item ``Bayes Rule tells us that we then take those prior odds and multiply them by the likelihood ratio of the blood/DNA evidence in order to arrive at the posterior odds in favour of the defendant’s paternity. The Court then has to consider whether those odds meet the required standard of proof. Thus the expert should say ‘however likely you think it is that the defendant is the father on the basis of the other evidence, my evidence multiplies the odds X times’." \parencite{robertson1992unhelpful}

    \item ``The main focus of attention will be confined to the perspective of how one can assess the value of scientific findings in order to inform about how findings should affect the views of others on selected issues in a case.'' ( \cite{biedermann2014liberties}, p. 182)
    
    \item ``For example, for a likelihood ratio of a thousand, the scientist may think of reporting along the following lines: ‘My findings are on the order of one thousand times more probable if the person of interest is the author of the questioned text than if an unknown person wrote the questioned text. Hence, whatever odds the recipient of expert information assesses that the person of interest is the author, based on other evidence, my findings multiply those odds by one thousand. For example, if the prior odds are even, then the posterior odds are one thousand, but will be less for smaller prior odds.'' \parencite{biedermann2018analysing}
    
    %\item ``Determination of the BF (Bayes' factor) is typically considered to be in the domain of the forensic scientist.'' \parencite{taroni2016dismissal}

\item In criminal adjudication, the values of the prior odds and the posterior odds are matters for the judge and jury, in accordance with the normal division of labour in forensic fact-finding. The value of the likelihood ratio, however, is a matter for the forensic scientist or other expert witness, as it is an assessment of the objective probative value of their evidence. Assessments of prior and posterior odds require subjective opinions which are the responsibility of the fact-finders. The scientist does not need to know values for either the prior or the posterior odds. The likelihood ratio, or a range of such ratios, can be calculated on the basis of the assumed truth of the propositions put forward by the prosecution and defence.   \parencite{aitken2008fundamentals} 

\item Part of the task of expert witnesses should be to explain how the court is helped by the evidence given. Why should the witness not suggest by precisely how much it should help the court? The witness could say something like: ‘Whatever the odds of the hypothesis versus the alternative based upon the other evidence (which I have not heard), my evidence makes them R times higher’, where R is the value of the likelihood ratio. This not only gives the correct value for the evidence but tells the jury what to do with it, whereas it is not self-evident what is to be done with a likelihood ratio.  (\cite{robertson2016interpreting}, p. 67)
\end{itemize}
 
These viewpoints espouse precisely the application of equation (5) in evidence communication, in which an expert is expected to provide the value of a likelihood ratio (or Bayes' factor) that someone else ``should'' use when applying Bayes' rule rather than emphasizing the need for each recipient  to individually assess the value of the opinion provided by the forensic expert.  

We have also noted instances where authors describe Bayes' rule and an expert providing a likelihood ratio without mentioning a recipient assigning their own weight to an expert's likelihood ratio.  We view such presentations as indirectly and perhaps inadvertently supporting the ``deferential-Bayes" equation by omission. We list some examples below, with added italics to highlight the most relevant phrases:

\begin{itemize}
    \item ``Bayes Theorem shows us that, while the investigator or court is concerned with questions of the type: `what is the probability that the suspect was at the crime scene?', \emph{the scientist, through the likelihood ratio, should address questions of the type `what is the probability of the evidence given that the suspect was not at the crime scene?'}\,'' \parencite{Evett87}
    
    \item ``The formula can be expressed in words as follows: Posterior odds = Likelihood ratio $\times$ Prior odds.  The court is concerned with questions of the kind ‘what is the probability that the defendant committed the crime given the evidence?’ but \emph{Bayes theorem demonstrates that, for the scientist to assist the court in updating its probabilities s/he must address questions of the kind ‘what is the probability of the evidence given that the defendant committed the crime?’}\,'' \parencite{Evett1998TowardsAU}

\begin{comment}	
\item ``The Bayesian model represents the application of probability theory to reasoning under uncertainty. The model reveals the central importance of the likelihood ratio, the formulation of which crystallises three key principles for the interpretation of forensic science evidence 	\parencite{evett2000impact}.
	\begin{description}
	\item 1. Interpretation of scientific evidence is carried out within a framework of circumstances. The interpretation depends on the structure and content of the framework. 
	\item 2. Interpretation is only meaningful when two or more competing propositions are addressed.  
	\item 3. \emph{The role of the forensic scientist is to consider the probability of the evidence given the propositions that are addressed.}'' 
	\end{description}
\end{comment}

    \item  ``Bayes’ Theorem provides a model that clearly distinguishes the role of the scientist and that of the fact finders. \emph{ The role of the scientist is to advise the fact finders on the strength of the evidence by assigning the $LR$. Any consideration of the prior or posterior odds (or the probability) of the propositions is left to the fact finders.}" \parencite{Buckleton_Stiffleman}
    
    \item  ``\emph{The role of the forensic scientist is to assign the probabilities of the evidence given the propositions that are considered.}'' \parencite{Buckleton_Stiffleman}
    
\end{itemize}

\section*{Appendix-C}
\noindent{\large \bf Bayesian Reasoning Applied to an Expert's Categorical Conclusion}

\begin{figure}
    \centering
\includegraphics[width=.9\linewidth]{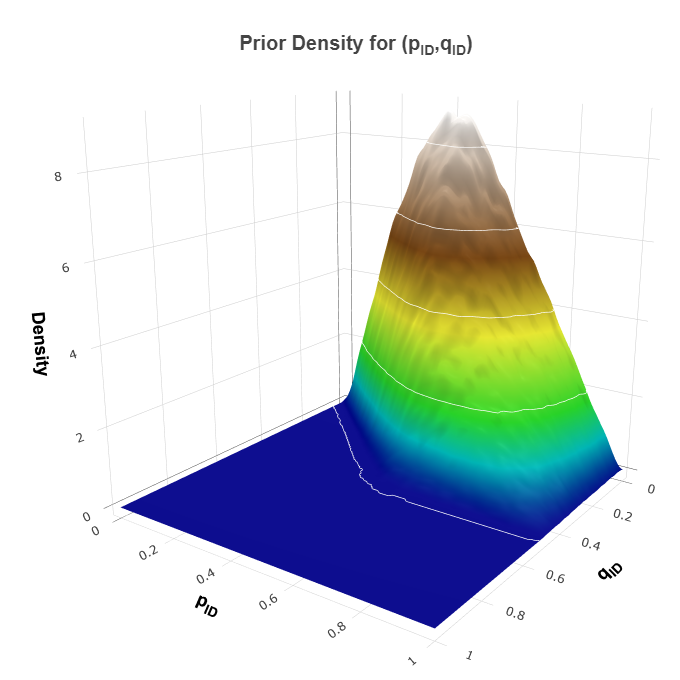}
    \caption{Joint prior density for (\pid,\qid).}
    \label{fig:fig01}
\end{figure}

\begin{figure}
    \centering
\includegraphics[width=.9\linewidth]{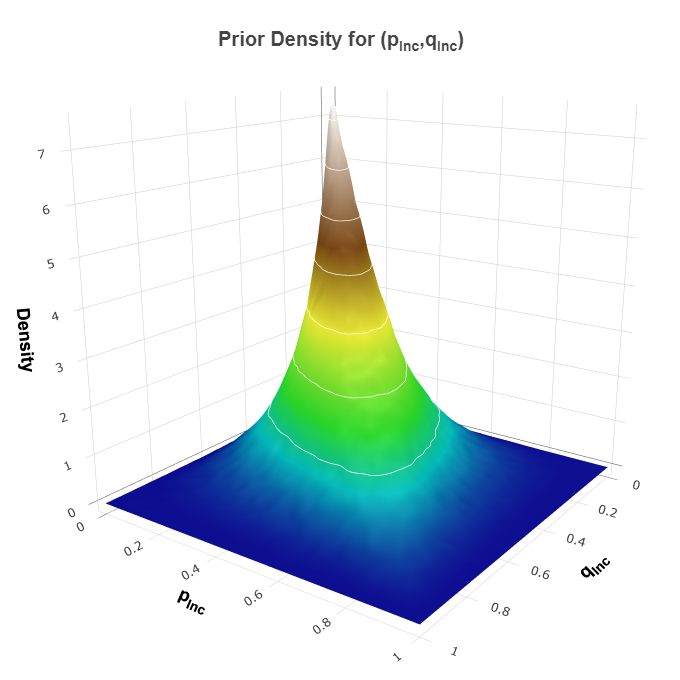}
    \caption{Joint prior density for (\pinc,\qinc).}
    \label{fig:fig02}
\end{figure}

\begin{figure}
    \centering
\includegraphics[width=.9\linewidth]{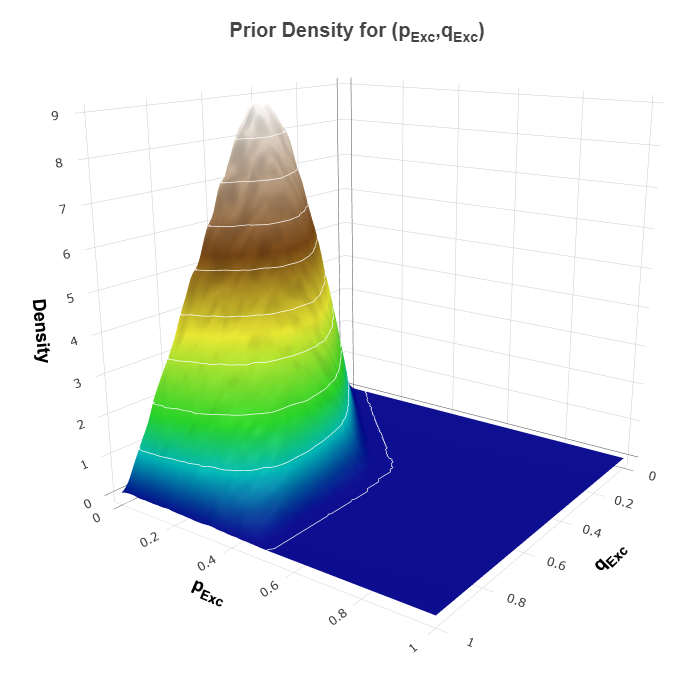}
    \caption{Joint prior density for (\pex,\qex).}
    \label{fig:fig03}
\end{figure}

This section provides additional details for the example originally described in Section~3.1.  Alice's initial uncertainty regarding how frequently Bert offers various conclusions under $H_1$ and $H_2$, respectively, is uniform over the sample space that satisfies her constraints.  We  learn about properties of this distribution using rejection sampling. In particular, we draw samples for (\pid, \pinc, \pex) and for (\qid, \qinc, \qex) from a pair of Dirichlet distributions, each with concentration parameters (1, 1, 1).  If the resulting six-element vector (\pid, \pinc, \pex, \qid, \qinc, \qex)  satisfies Alice's constraints, we keep it.  If not, we discard it.  We repeat this process until we have a million draws from her distribution.  Figure~\ref{fig:fig01}, Figure~\ref{fig:fig02}, Figure~\ref{fig:fig03} display the joint prior densities for (\pid,\qid), (\pinc,\qinc), and (\pex,\qex), respectively. 

To compute Alice's  $LR$ for Bert concluding ``ID'', we take the average of \pid\ among the million accepted draws and divide by the average of \qid.  Analogous steps provide Alice's $LR$ for inconclusive and exclusion conclusions.

\begin{figure}
    \centering
\includegraphics[width=.9\linewidth]{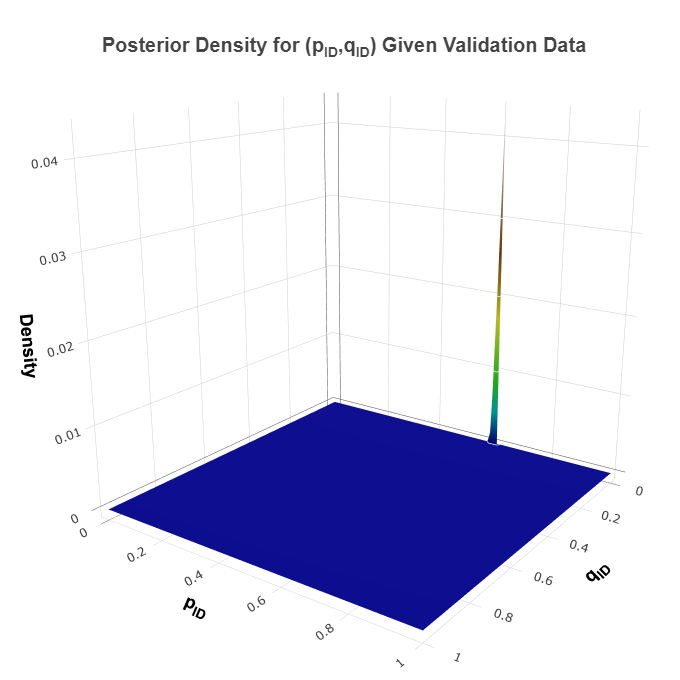}
    \caption{Joint posterior  density for (\pid,\qid) after taking into account  validation data.}
    \label{fig:fig1}
\end{figure}

\begin{figure}
    \centering
\includegraphics[width=.9\linewidth]{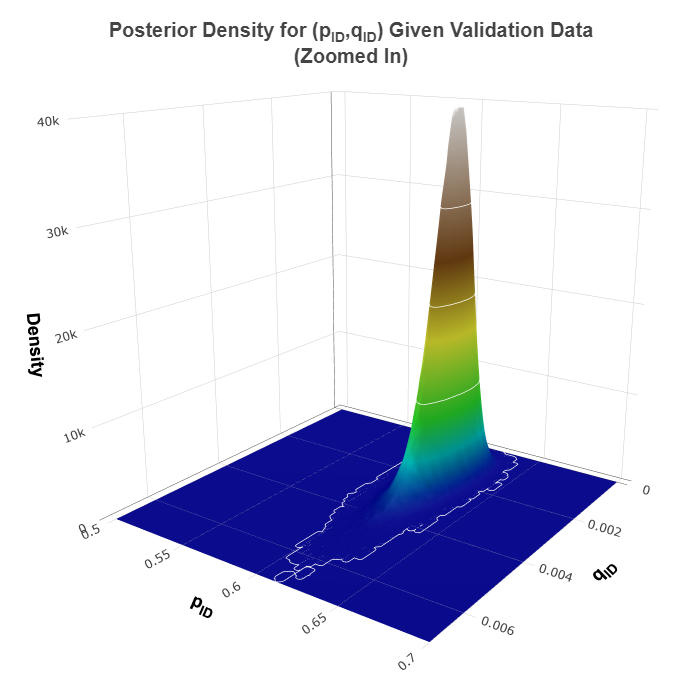}
    \caption{Zoomed in view of the joint posterior  density for (\pid,\qid) after taking into account  validation data.}
    \label{fig:fig1a}
\end{figure}

\begin{figure}
    \centering
\includegraphics[width=.9\linewidth]{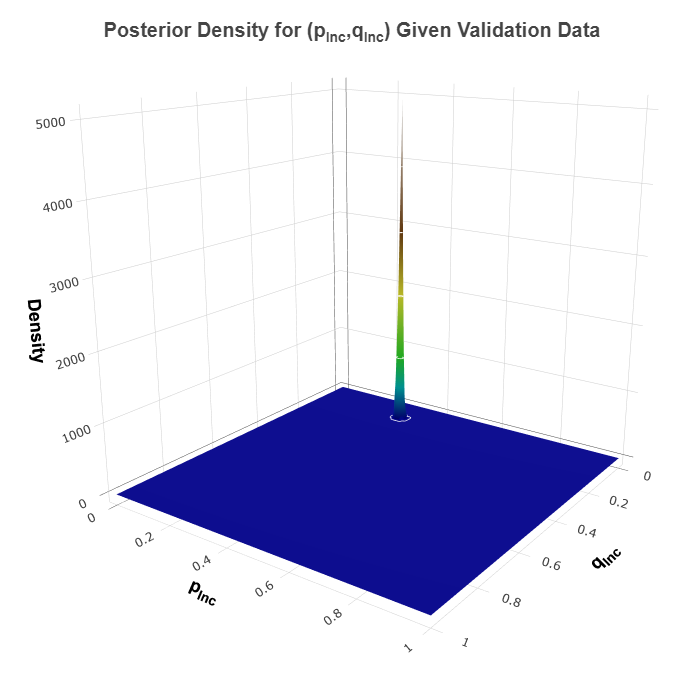}
    \caption{Joint posterior  density for (\pinc,\qinc) after taking into account  validation data.}
    \label{fig:fig2}
\end{figure}

\begin{figure}
    \centering
\includegraphics[width=.9\linewidth]{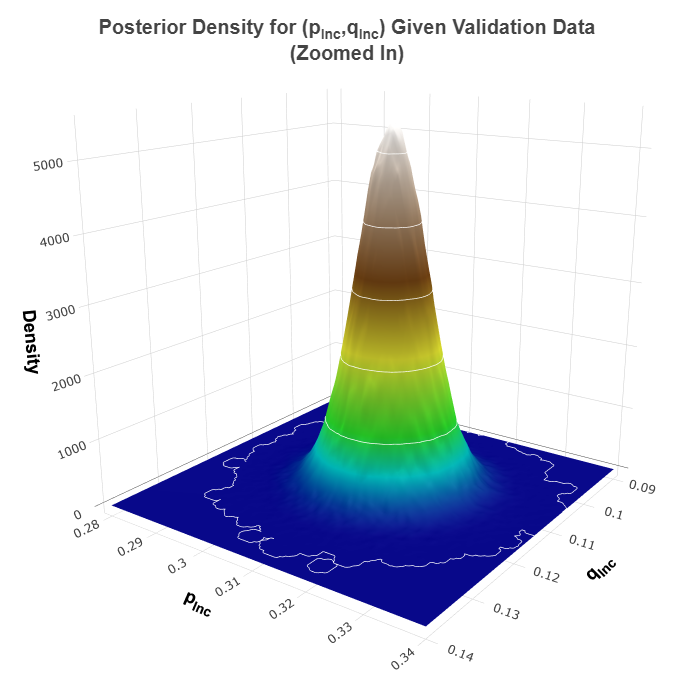}
    \caption{Zoomed in view of the joint posterior  density for (\pinc,\qinc) after taking into account  validation data.}
    \label{fig:fig2a}
\end{figure}

\begin{figure}
    \centering
\includegraphics[width=.9\linewidth]{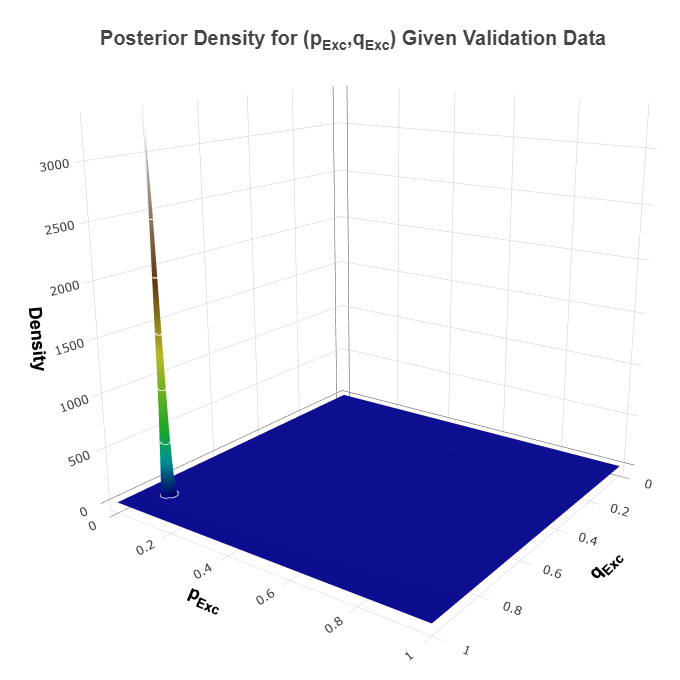}
    \caption{Joint posterior density for (\pex,\qex) after taking into account  validation data.}
    \label{fig:fig3}
\end{figure}

\begin{figure}
    \centering
\includegraphics[width=.9\linewidth]{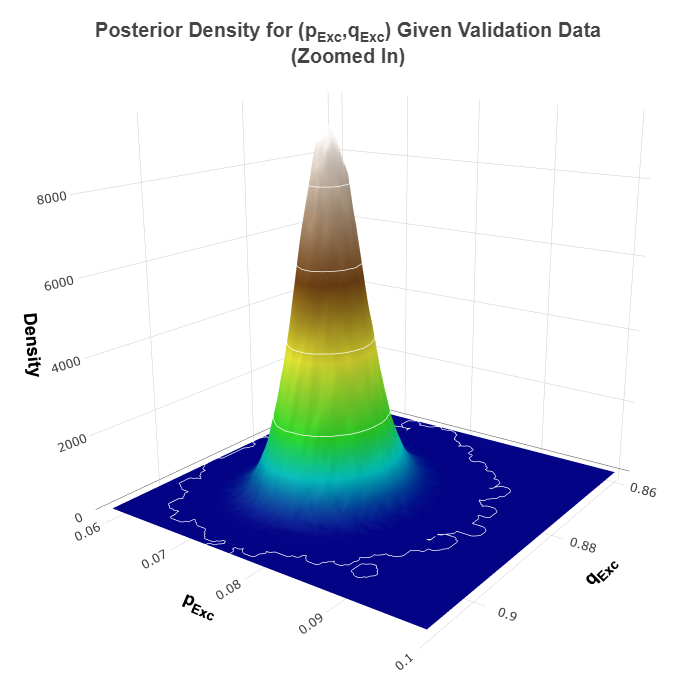}
    \caption{Zoomed in view of the joint posterior  density for (\pex,\qex) after taking into account  validation data.}
    \label{fig:fig3a}
\end{figure}

To compute Alice's $LR$ after receiving validation data, we use the fact that a Dirichlet distribution is the conjugate prior \parencite{DeGroot_2004} for the probability vector in a multinomial distribution.  This means we can sample from Alice's posterior distribution using rejection sampling with a Dirichlet distribution as with Alice's prior.  The updated Dirichlet distribution for (\pid, \pinc, \pex) has concentration parameters ($n_{ID_1}+1, n_{Inc_1}+1, n_{Exc_1}+1)$, where $n_{ID_1}, n_{Inc_1},$ and $n_{Exc_1}$ are the number of ID, inconclusive, and exclusion conclusions occurring among scenarios representative of $H_1$ in the validation tests. Similarly, the updated Dirichlet distribution for (\qid, \qinc, \qex) has concentration parameters ($n_{ID_2}+1, n_{Inc_2}+1, n_{Exc_2}+1)$, where $n_{ID_2}, n_{Inc_2},$ and $n_{Exc_2}$ are the number of ID, inconclusive, and exclusion conclusions occurring among scenarios representative of $H_2$ in the validation tests.
Figure~\ref{fig:fig1}, Figure~\ref{fig:fig2}, Figure~\ref{fig:fig3} display Alice's joint posterior densities for (\pid,\qid), (\pinc,\qinc), (\pex,\qex), respectively, after she examines the validation data. 
Figure~\ref{fig:fig1a}, Figure~\ref{fig:fig2a}, Figure~\ref{fig:fig3a} display zoomed-in versions of Figure~\ref{fig:fig1}, Figure~\ref{fig:fig2}, Figure~\ref{fig:fig3}, respectively.

\subsection*{Applying Bayesian Reasoning to an Expert's $LR$}

This section provides additional details for the example originally described in Section~3.2. After comparing the impressions, Bert summarizes his findings to Alice by specifying the propositions he considered and stating his weight of evidence is $r$ (i.e., $LR_B=10^r$). Alice now seeks to assign an $LR$ for this new information, which requires her to assess the likelihoods that Bert would say $log_{10}(LR_B) = r$ under $H_1$ and  $H_2$, respectively. 

Alice considers her uncertainty regarding what distributions would reflect the weights of evidence Bert would articulate when comparing two impressions from the same finger (corresponding to $H_1$) or when analyzing two impressions from different fingers (corresponding to $H_2$). She assumes that $log_{10}(LR_B)$ would be normally distributed for $H_1$ and $H_2$, respectively, but is uncertain about the mean and variance for each of these two distributions.  Alice conveys her uncertainty in the parameters $(\mu_1,\sigma_1^2)$ and $(\mu_2,\sigma_2^2)$ using normal-gamma distributions, which are the conjugate priors for normal distributions with unknown means and variances  (page 268, \cite{bernardo2009Bayesian}), meaning it is computationally simple to update these priors based on new information from examiner performance.  That is, Alice assumes  
\begin{align*}
log_{10}(LR_B)|H_1,\mu_1,\sigma_1^2,  \mu_2, \sigma_2^2 &\sim \mbox{Normal}(\mu_1,\sigma_1^2) \\ \mbox{~~and~~}log_{10}(LR_B)|H_2, \mu_1,\sigma_1^2, \mu_2, \sigma_2^2 &\sim \mbox{Normal}(\mu_2,\sigma_2^2).
\end{align*}
To simplify notation, precision (i.e., reciprocal of  variance) is used in place of variance. That is, we use $\tau_1 = 1/\sigma_1^2$ and $\tau_2 = 1/\sigma_2^2$.

Alice's prior for $(\mu_1,\tau_1)$, which is given according to 
$$
(\mu_1,\tau_1) \sim \mbox{Normal-Gamma} (\mu_{10},n_{\mu 1},\tau_{10},n_{\tau 1}), 
$$
is specified in terms a prior mean of $\mu_{10} = 5$ with $n_{\mu 1 } = 1$ observation's worth of information about the mean and a prior precision of $\tau _{10} = \myfrac{1}{100}$ with $n_{\tau 1} = 1$ observation's worth of information about the precision.  Similarly, her prior for $(\mu_2,\tau_2)$, which is given according to 
$$
(\mu_2,\tau_2) \sim \mbox{Normal-Gamma} (\mu_{20},n_{\mu 2},\tau_{20},n_{\tau 2}),
$$
is specified as having a prior mean of $\mu_{20} = -5$ with $n_{\mu 2 } = 1$ observation's worth of information about the mean and a prior precision of $\tau _{20} = \myfrac{1}{100}$ with $n_{\tau 2} = 1$ observation's worth of information about the precision.  Finally, Alice assumes the pair $(\mu_1,\tau_1)$ to be independent of the pair $(\mu_2,\tau_2)$.

Suppose Bert, in addition to stating his weight of evidence for the case at hand, also provides results from validation testing where he was asked to evaluate $LR$s in reference scenarios where a third party knew whether or not the impressions being evaluated were from the same source.  Suppose among the provided results, there were $n_1$ tests that Alice views as having come from the same distribution as $LR_B$ would have if $H_1$ were true, and that the logarithms of these $LR$s have a sample mean of $\bar{y}_1$ and a sample variance of $s_1^2$.

Learning about these validation test results reduces Alice's uncertainty regarding the distribution of $log_{10}(LR_B)$ values under $H_1$.   In particular, as shown in \verb@https://en.wikipedia.org@\\\verb@/wiki/Normal-gamma_distribution@, an application of Bayes' rule reveals that her updated uncertainty would follow a normal-gamma distribution with parameters:
\begin{align}
\mu^*_1 &=\myfrac{n_{\mu 1}\mu_1+n_1\bar{y}_1}{n_{\mu 1}+n_1},\\
n^*_{\mu 1}&=n_{\mu 1}+n_1\\ 
n^*_{\tau 1}&=n_{\tau 1}+n_1, \mbox{~~and}\\ 
\myfrac{n^*_{\tau 1}}{\tau^*_1}&=\myfrac{n_{\tau 1}}{\tau_1}+n_1 s^2_1+\myfrac{n_{\mu 1} n_1 (\bar{y}_1-\mu_1)^2}{n_{\mu 1}+n_1}
\end{align}

\subsection*{Applying Bayesian Reasoning to an Expert's $LR$ Range}

Here we give the computational details for the example discussed in Section~3.3 in which Bert  provided a range for a likelihood ratio of 100 million to 10 billion.  
To illustrate one way of applying Bayesian reasoning in this scenario, we can directly build on the example from Section~3.2  by converting the end points of the provided interval to a $log_{10}$ scale and taking the interval midpoint $m$ and interval width $w$. For example, the interval of 100 million to 10 billion would be converted to a $log_{10}$ scale as 8 to 10.  The midpoint $m$ would be 9 and the interval width $w$ would be 2.  Suppose Alice forms a prior distribution for the distribution of $(m,w)$ in the following manner.
Alice treats $m$ exactly as $log_{10}(LR_B)$ is treated in the example in which Bert provided a single $LR$ value.  Further, Alice assumes $w$ is independent of $m$, both under $H_1$ and under $H_2$.  Under this setup, Alice's $LR$ for an interval provided by Bert, $Interval_B$, can be decomposed as follows:
\begin{align*}
   LR_A(Interval_B)&=LR_A(m,w)\\
%   &=\myfrac{Pr(m,w|H_1)}{Pr(m,w|H_2)} \\
%&=\myfrac{Pr(m|H_1)Pr(w|H_1)}{Pr(m|H_2)Pr(w|H_2)} \\
%&=\myfrac{Pr(m|H_1)}{Pr(m|H_2)}\myfrac{Pr(w|H_1)}{Pr(w|H_2)} \\
&=LR_A(m)LR_A(w)
\end{align*}
\noindent $LR_A(m)$ behaves identically as in the example from Section~3.2, so we turn our attention to $LR_A(w)$.  Suppose Alice assumes that $w$ follows a gamma distribution for $H_1$ and $H_2$, respectively,  but is uncertain about the value of the parameters for shape ($\alpha$) and rate ($\beta$) for each of these two distributions.  That is, Alice assumes
\begin{align*}
w|H_1,m,\alpha_1,\beta_1,\alpha_2,\beta_2 &\sim gamma(\alpha_1,\beta_1) \\ \mbox{~~and~~}w|H_2,m,\alpha_1,\beta_1,\alpha_2,\beta_2 &\sim gamma(\alpha_2,\beta_2).
\end{align*}
Suppose Alice chooses to convey her uncertainty in the parameters $(\alpha_1,\beta_1)$ and $(\alpha_2,\beta_2)$ using conjugate priors for gamma distributions with unknown shape and rate parameters.  The density of the conjugate distribution in this case is proportional to $\myfrac{p^{\alpha-1}e^{-\beta q}}{\Gamma(\alpha)^r\beta^{-\alpha s}}$ (\cite{miller1980}; see also \textcite{Wiki})
%\verb|https://en.wikipedia.org/wiki/Conjugate_prior|). 

Suppose Alice's prior for $(\alpha_1,\beta_1)$ is specified with hyperparameters $p=9,q=6,r=2$ and $s=2$.  The specification of $r=2$ and $p=9$ can be interpreted as having two observation's worth of information about $\alpha$ and that the product of those observations is 9.  The specification of $s=2$ and $q=6$ can be interpreted as having two observation's worth of information about $\beta$ and that the sum of those observations is 6.   

While one might expect a recipient to have at least a slight inclination that $H_1$ scenarios will tend to produce higher interval mid-points than $H_2$, we do not see an obvious reason to pick either $H_1$ or $H_2$ as expected to lead to greater interval widths than the other. Suppose that Alice  therefore uses the same values of $p=9,q=6,r=2$ and $s=2$ to characterize her uncertainty in $(\alpha_2,\beta_2)$ as she did for $(\alpha_1,\beta_1)$.  

\begin{figure}
    \centering
\includegraphics[width=.9\linewidth]{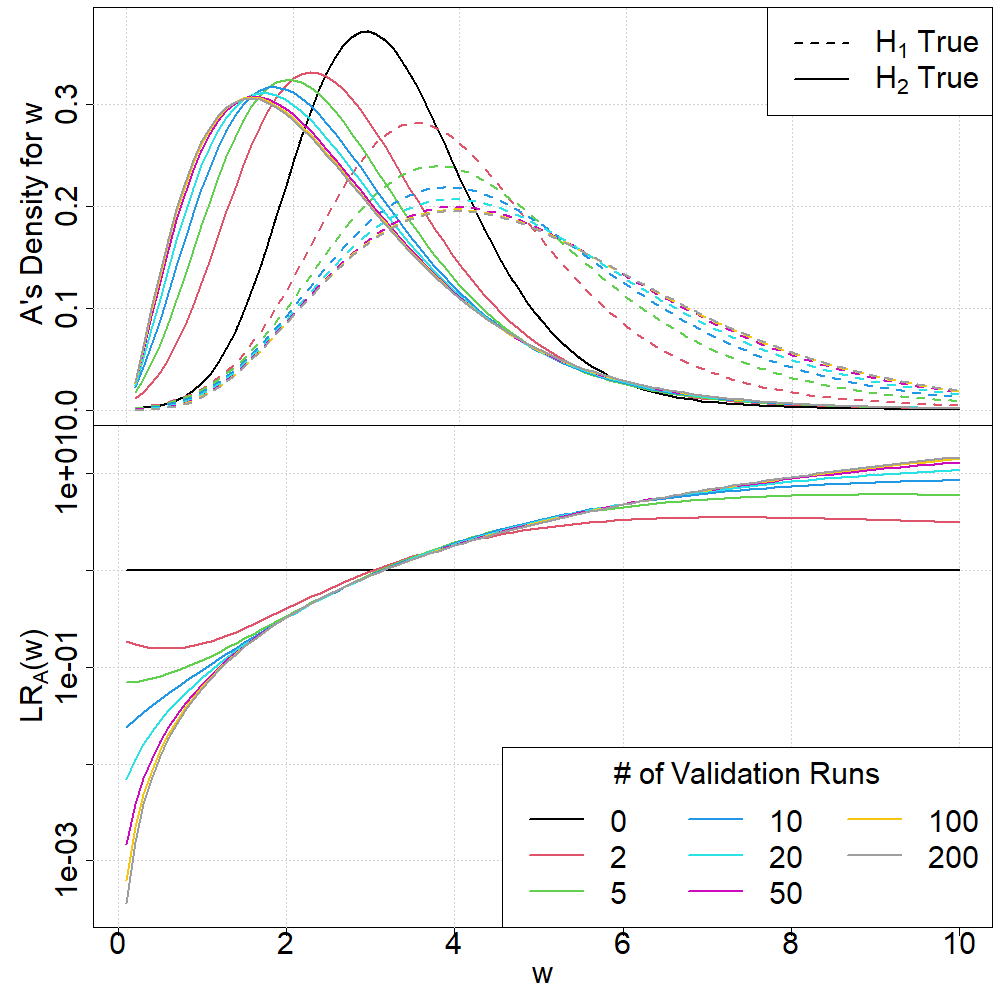}
    \caption{Top: Distributions for $w$, the width of the interval provided for $log_{10}(LR_B)$, under $H_1$ (dashed curves) and under $H_2$ (solid curves) for varying numbers of validation tests. Bottom: Value of $LR_A (w)$ as a function of $w$ after disclosure of results from varying numbers of validation tests.}
    \label{fig:LR_w_post}
\end{figure}
The black curve in the top panel of Figure \ref{fig:LR_w_post} shows the marginal distribution for $w$ under the assumed priors. As seen in the black line in the bottom panel, $LR_A(w)=1$ for all values of $w$ because the marginal distribution for $w$ is the same for both $H_1$ and $H_2$ in this setup. Thus, the value of $w$ from a case interpretation would not influence the recipient's posterior probability for $H_1$ unless performance data is also provided. Of course, if values of $w$ provided from scenarios for $H_1$ had a similar sample size, product, and sum as values provided from scenarios for $H_2$, then $LR_A(w)$ would also stay close to one.  In order for $LR_A(w)$ to move away from one, Alice would need to receive information showing differences between what values of $w$ have occurred under scenarios representing $H_1$ and under scenarios representing $H_2$.  For instance, suppose Alice is presented with $n$ observations of $w$ from $H_1$ scenarios that have a product of $4.5^{n}$ and a sum of $5n$.  Suppose Alice also receives $n$ observations of $w$ from $H_2$ scenarios that have a product of $2^{n}$ and a sum of $2.5n$.  Figure~\ref{fig:LR_w_post} shows the posterior marginal distribution for $w$ under $H_1$ and $H_2$, respectively, as well as the corresponding $LR_A(w)$ profile for different values of $n$.

There is no mystery about how a Bayesian recipient should rationally update their prior odds into posterior odds based on an interval provided by an expert.  Regardless of the form of information an expert provides, a Bayesian recipient processes that information by following the same general steps of evaluating a likelihood for that new information under each proposition of interest to them and applying Bayes' rule.  However,  there is a practical question of whether an interval is more helpful to the recipient than a single scalar value. As discussed in Section~\ref{sec:bayesianrecipients}, the answer would depend on how strongly available data demonstrates a high level of discrimination for the current result among propositions of likely interest, compared to the demonstrated discrimination for the scalar.

\subsection*{Applying Bayesian Reasoning with Two Experts}

Here we consider the situation introduced in Section~3.4 where Alice received opinions from two experts, Bert and Carla. %We recognize that fully absorbing this example requires some familiarity with multivariate statistics, which many readers may not have.  The main body of the paper does not require a deep understanding of this example, which is why it has been moved to an appendix.
%Suppose a recipient (identified using subscript $A$) hears from two experts, denoted as $B_p$ and $B_d$, who were respectively called by the prosecution and defense.  
They both evaluated the same piece of evidence with respect to the same two propositions, say $H_1$ and $H_2$, and provide their respective $LR$s, say $LR_B$ and $LR_C$, to Alice.  We illustrate how Alice can apply Bayesian reasoning in response to this new information.  For simplicity, we suppose Alice is interested in the same two propositions as were considered by the experts, namely $H_1$ and $H_2$.  As with the previous examples, Alice will assess how likely it would be to encounter the new information under each of the propositions of interest to her.  This can be accomplished by specifying a distribution for the pair of expert $LR$s under $H_1$ and $H_2$.

Suppose Alice assumes that, for both $H_1$ and $H_2$, the pair $\left(log_{10}(LR_B),log_{10}(LR_C)\right)$ follows a bivariate normal distribution, which has parameters $\bm{\mu}$ (a two-element vector representing the average $log_{10}(LR)$ from each of the two experts) and $\bm{\Sigma}$ (a two-by-two covariance matrix that reflects the variability of each expert's $log_{10}(LR)$s across cases and the correlation of the $log_{10}(LR)$s between the two experts).  Alice expects the experts' behaviors to differ between scenarios representing $H_1$ and $H_2$ but is uncertain about the mean vector and covariance matrix that would reasonably reflect the experts' behaviors under either scenarios. For computational simplicity, we reflect Alice's   uncertainty using conjugate priors for this scenario.  In particular, for $H_1$ suppose Alice uses a Wishart($\Lambda_0,n_0=2$) distribution with prior precision matrix $\Lambda_0 = \bigl( \begin{smallmatrix}0.2 & -0.15\\ -0.15 & 0.2\end{smallmatrix}\bigr)$ to specify uncertainty regarding the covariance matrix of the experts' $log_{10}(LR)$s under $H_1$ scenarios. Here $n_0=2$ stands for two degrees of freedom or two observations' worth of information regarding this covariance matrix. Alice further specifies a prior mean vector of $\mu_{1,0}=\bigl( \begin{smallmatrix}4\\ 2\end{smallmatrix}\bigr)$ with $k_0=2$ observations' worth of prior information regarding this mean vector.

This gives 
\[
\Lambda \sim \mbox{Wishart}(\Lambda_0=\bigl(\begin{smallmatrix}0.1 & -0.08\\ -0.08 & 0.1\end{smallmatrix}\bigr),n_0=2)
\] and
\[
\mu|\Lambda,k_0 \sim Normal\bigl(\mu_{1,0}=\bigl( \begin{smallmatrix}5\\ 5\end{smallmatrix}\bigr),\Sigma = \bigl(2 \Lambda\bigr)^{-1} \bigr).
\]

The marginal distribution of $\left(log_{10}(LR_{B_p}),log_{10}(LR_{B_d})\right)^T$ is a bivariate Student-t distribution with the following parameters: $n_0$ degrees of freedom; mean vector/non-centrality parameters $ \mu_{1,0}$; and scale matrix $(\frac{k_0(n_0-1)}{k_0+1}\Lambda_0)^{-1}$.

Suppose Alice uses the same parameters to reflect uncertainty in the distribution of $\left(log_{10}(LR_B),log_{10}(LR_C)\right)^T$ under $H_2$ as were used under $H_1$, with the exception that $\mu_{2,0}=\bigl( \begin{smallmatrix}-2\\ -4\end{smallmatrix}\bigr)$ is used in place of $\mu_{1,0}$.

Suppose Bert and Carla provide $LR_B=100$ and $LR_B=30$, respectively, for the case at hand. The priors specified above produce $LR_A$ of 4.35.  That is, in  the absence of any performance data, Alice is less persuaded by the experts' opinions of the evidence than either of the experts were of the evidence itself. 

To illustrate the effect of performance data on Alice, suppose Alice also receives performance data showing what $LR$s the experts have provided in response to $m_1$ scenarios reflecting $H_1$. Assume both experts have declared their respective $LR$s in each of these scenarios.  Denote these data from the $H_1$ scenarios as $\mathbf{x}_{1,1},\mathbf{x}_{1,2},\ldots,\mathbf{x}_{1,m_1}$, where $\mathbf{x}_{1,i} = 
\bigl( \begin{smallmatrix} log_{10}(LR_B,1,i)\\ 
log_{10}(LR_C,1,i)
\end{smallmatrix}
\bigr)$ 
is the pair of $log_{10}(LR)$ values provided by the experts in the $i^{\mbox{\footnotesize th}}$ sample under $H_1$.  
Suppose the sample average for these data is $\bar{\mathbf{x}}_1=\bigl( \begin{smallmatrix}3.5\\ 2.5\end{smallmatrix}\bigr)$ and the sample scatter matrix is given by $S_1=m_1  \bigl( \begin{smallmatrix}5 & 4\\ 4 & 5\end{smallmatrix}\bigr)$.  Given this new information about the experts' behavior, Alice updates her prior for the distribution of $\bigl( \begin{smallmatrix}log_{10}(LR_B)\\ log_{10}(LR_C)\end{smallmatrix}\bigr)$ under $H_1$ according to:
\begin{align*}
n_1&=n_0+m_1\\
k_1&=k_0+m_1\\
\mu_1&=\frac{k_0\mu_{1,0}+m_1\bar{\mathbf{x}}_1}{k_0+m_1}\\
\Lambda_1&= \bigl( \Lambda_0^{-1}+S_1+\frac{k_0m_1}{k_0+m_1}(\bar{\mathbf{x}}_1-\mu_{1,0})(\bar{\mathbf{x}}_1-\mu_{1,0} )^T\bigr)^{-1}
\end{align*}
Similarly, for $H_2$ suppose that $m_2$ samples have a sample average of $\bar{\mathbf{x}}_2=\bigl( \begin{smallmatrix}-2.5\\ -3.5\end{smallmatrix}\bigr)$ and a sample scatter matrix  given by $S_2=m_2  \bigl( \begin{smallmatrix}5 & 4\\ 4 & 5\end{smallmatrix}\bigr)$.  Given this new information about the experts' behavior in scenarios reflecting $H_2$, the recipient updates their prior  for the distribution of $\bigl( \begin{smallmatrix}log_{10}(LR_B)\\ log_{10}(LR_C)\end{smallmatrix}\bigr)$ under $H_2$ according to:
\begin{align*}
n_2&=n_0+m_2\\
k_2&=k_0+m_2\\
\mu_2&=\frac{k_0\mu_{2,0}+m_2\bar{\mathbf{x}}_2}{k_0+m_2}\\
\Lambda_2&= \bigl( \Lambda_0^{-1}+S+\frac{k_0m_2}{k_0+m_2}(\bar{\mathbf{x}}_2-\mu_{2,0})(\bar{\mathbf{x}}_2-\mu_{2,0} )^T\bigr)^{-1}
\end{align*}

\begin{figure}
    \centering
    \includegraphics[width=.7\linewidth]{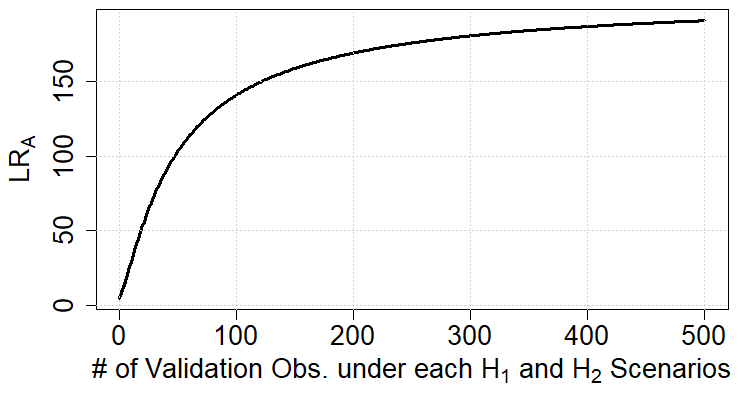}
    \caption{Curve showing $LR_A$ for a pair of expert $LR$s after Alice is provided with results of various numbers of tests performed under $H_1$-true scenarios and $H_2$-true scenarios.} 
    \label{fig:two_expert_LR}
\end{figure}

Figure \ref{fig:two_expert_LR} illustrates the effect of the validation data on $LR_A$ as a function of equal numbers of samples collected under $H_1$ and under $H_2$ (i.e., $m_1$=$m_2$).  As one might expect, learning about validation data that shows strong differences between the $LR$ values that the experts tend to provide under scenarios reflecting $H_1$ and $H_2$, respectively, strengthens the recipient's confidence in the experts and increases the weight the recipient gives to the experts' opinions. 

\newpage

\end{document}